\documentclass[dvips]{acta}
\usepackage{supertabular,lscape,epsfig}
\usepackage{amssymb}
\usepackage{amsmath}

\SetPages{1}{1000}

\SetVol{58}{2008}

\begin{document}

\newcommand{\sa}[1]{\{#1\}}
\newcommand{\sab}[1]{\bigg\{#1\bigg\}}
\newcommand{\ta}[1]{\big<#1\big>}
\newcommand{\tab}[1]{\Big<#1\Big>}
\newcommand{\tabb}[1]{\bigg<#1\bigg>}

\begin{Titlepage}

\Title{Frequency analysis of Cepheids in the Large Magellanic Cloud.\\
       New types of classical Cepheid pulsators}

\medskip

\Author{P.~~M~o~s~k~a~l~i~k$^1$ ~and~~ Z.~~K~o~{\l}~a~c~z~k~o~w~s~k~i$^{2,3}$}
{$^1$ Copernicus Astronomical Centre, ul.~Bartycka~18, 00--716 Warsaw, Poland
\\
e-mail: pam@camk.edu.pl
\\
$^2$ Universidad de Concepci\'on, Departamento de Fisica, Casilla 160--C, Concepci\'on, Chile
\\
e-mail: zibi@astro-udec.cl
\\
$^3$ Astronomical Institute, University of Wroc{\l}aw, ul.~Kopernika~11, 51--622 Wroc{\l}aw, Poland}

\medskip

\Received{September 1, 2008}

\end{Titlepage}

\Abstract{We have performed a detailed systematic search for
multiperiodicity in the Population~I Cepheids of the Large
Magellanic Cloud. In this process we have identified for the first
time several new types of Cepheid pulsational behaviour. We have
found two triple-mode Cepheids pulsating simultaneously in the
first three radial overtones. In 9\% of the first overtone
Cepheids we have detected weak, but well resolved secondary
periodicities. They appear either very close to the primary
pulsation frequency or at a much higher frequency with a
characteristic period ratio of $0.60-0.64$. In either case, the
secondary periodicities must correspond to nonradial modes of
oscillation. This result presents a major challenge to the theory
of stellar pulsations, which predicts that such modes should not
be exited in Cepheid variables. Nonradial modes have also been
found in three of the fundamental/first overtone double-mode
Cepheids, but no such oscillations have been detected in single
mode Cepheids pulsating in the fundamental mode.

In 19\% of double-mode Cepheids pulsating in the first two radial
overtones (FO/SO type) we have detected a Blazhko-type periodic
modulation of amplitudes and phases. Both modes are modulated with
a common period, which is always longer than 700\thinspace days.
Variations of the two amplitudes are anticorrelated and maximum of
one amplitude always coincides with minimum of the other. We have
compared observations of modulated FO/SO Cepheids with predictions
of theoretical models of the Blazhko effect, showing that
currently most popular models cannot account for  properties of
these stars. We propose that Blazhko effect in FO/SO Cepheids can
be explained by a nonstationary resonant interaction of one of the
radial modes with another, perhaps nonradial, mode of
oscillations.
}
{Stars: variables: Cepheids -- stars: oscillations: nonradial --
stars: oscillations: multimode -- stars: oscillations: Blazhko
effect -- galaxies: Magellanic Clouds}

\section{Introduction}%1

Classical Cepheids and RR~Lyrae stars are among the best studied
pulsating variables. For decades they have been considered primary
example of purely radial pulsators, displaying large amplitude
oscillations, with only one or two radial modes excited. This
simple picture was challenged in the end of the last century when
Olech {\it et~al.}\thinspace (1999) discovered low amplitude
nonradial modes in three RR~Lyr stars of globular cluster M55. In
following years it was shown that presence of secondary
periodicities identified with nonradial modes of oscillations is a
very common property of RR~Lyr variables. They were detected in a
substantial fraction of stars in every studied RR~Lyr population:
in LMC (Alcock {\it et~al.}\thinspace 2000, 2003; Soszy\'nski {\it
et~al.}\thinspace 2003; Nagy \& Kov\'acs 2006), SMC (Soszy\'nski
{\it et~al.}\thinspace 2002), Galactic Bulge (Moskalik \& Poretti
2002, 2003; Mizerski 2003; Collinge {\it et~al.}\thinspace 2006),
Galactic field (Wils {\it et~al.}\thinspace 2006; Szczygie{\l} \&
Fabrycky 2007, see also Smith 1995) and most recently in
$\omega$~Cen (Olech \& Moskalik 2008). Depending on the stellar
system, nonradial modes are excited in between 5\% and 30\% of all
RR~Lyr stars.

Motivated by these findings we decided to search for nonradial
modes in classical (Population~I) Cepheids. As far as pulsation
properties are concerned, Cepheids are very close siblings of
RR~Lyr stars. It is therefore very interesting to check if
nonradial modes are excited in these pulsators as well. With this
goal in mind, we conducted a systematic search for
multiperiodicity among Pop.\thinspace I Cepheids of the Large
Magellanic Cloud (LMC). The primary source of data for this
analysis was the OGLE-II photometry (\.Zebru\'n {\it
et~al.}\thinspace 2001). The data, collected du\-ring microlensing
survey, covers the timebase of 1000-1200\thinspace days, with
250-500 I-band measurements per star. Such a long, uniform and
high quality dataset is particularly well suited for our purpose.
Partial results of this survey were published by Moskalik,
Ko{\l}aczkowski \& Mizerski (2004, 2006) and by Moskalik \&
Ko{\l}aczkowki (2008a,b). In this paper we present a full
discussion of our findings and provide a complete inventory of
newly identified multiperiodic Cepheids in the LMC.

\section{Time series analysis of Cepheid lightcurves}%2

We analyzed all fundamental mode (FU), first overtone (FO) and
double-mode (FU/FO and FO/SO) LMC Cepheids listed in the OGLE-II
catalogs (Udalski {\it et~al.} 1999b; Soszy\'nski {\it
et~al.}\thinspace 2000), nearly 1300 stars in total. We omitted
all variables marked by the authors as FA or BR. The former are
most likely Pop.\thinspace II Cepheids, while the status of the
latter group is somewhat unclear. We used the I-band OGLE-II data
obtained with the Difference Image Analysis (or DIA) method. The
search for secondary periodicities was conducted with a standard
consecutive prewhitening technique. First, we fitted the data with
a Fourier sum describing pulsations with the dominant mode:

\begin{equation}
I(t) = \langle I\rangle + \sum_{k} A_{k} \sin (2\pi kf_1 t + \phi_{k}).
\end{equation}

\noindent The frequency of the mode, $f_1$, was also optimized. For
double-mode Cepheids, we fitted the data with double-frequency
Fourier sum, representing pulsations in two dominant (radial) modes:

\begin{equation}
I(t) = \langle I\rangle + \sum_{j,k} A_{jk} \sin [2\pi(jf_1 + kf_2)t + \phi_{jk}].
\end{equation}

\noindent The residuals of the fit were then searched for
secondary frequencies. This was done with the Fourier transform,
calculated over the range of $0-6$\thinspace c/d. If any new
periodicity was detected, a new Fourier sum, including {\it all
frequencies identified} so far, was fitted to the data and the fit
residuals were examined again. The process was stopped when no new
significant frequencies appeared. At this stage we performed data
clipping, rejecting all measurements deviating from the fitted
function by more than $5\sigma$, where $\sigma$ is the standard
deviation of the fit residuals. OGLE-II photometry is very clean
and only for a small number of stars 1--3 datapoints were removed
by this criterion. After data clipping, the frequency analysis was
repeated.

As a final step of analysis we checked all Cepheids with detected
secondary peaks for a possible lightcurve contamination. It is a
well known shortcoming of the DIA reduction method, that the
measured flux of a star can be contaminated by light coming from
neighbouring stars ({\it i.e.}\thinspace Mizerski \& Bejger 2002;
Hartman {\it et~al.}\thinspace 2004). As a result, some of the
discovered secondary frequencies might originate not in the
Cepheid, but in another variable star in Cepheid's neighbourhood.
In order to remove such spurious detections we proceeded in the
following way. First, we identified all variable stars from
OGLE-II catalog (\.Zebru\'n {\it et~al.}\thinspace 2001), which
are located within $r=1$\thinspace arcmin of the studied Cepheid.
For each of these stars we performed frequency analysis and then
checked if detected frequencies {\it or their aliases} appear as
secondary peaks in Cepheid's power spectrum. With this procedure
we weeded out 14 contaminated Cepheids. In LMC fields SC1 to SC12
we extended the search radius to $r=1.5$\thinspace arcmin, but no
new cases of contamination were found. The OGLE-II catalog of
variable stars is not complete, though. It misses most small
amplitude variables, which can contaminate lightcurves of nearby
stars as well. Therefore, we performed additional contamination
check, calculating frequency spectra of {\it all stars} (not only
known variables) within $r=20$\thinspace arcsec of the Cepheid. No
more contaminated Cepheids were rejected at this step.

\section{New double-mode Cepheids}%3

In the course of our analysis we identified five new "canonical"
double-mode Cepheids, pulsating in two radial modes. In four of them
the fundamental mode and the first overtone are excited (FU/FO
type), while in the fifth Cepheid the first and second overtones are
excited (FO/SO type). Basic properties of these variables are listed
in Table~1. Periods and amplitudes given in the table (as well as in
all other tables in this paper) are determined through the least
square fits of appropriate multifrequency solutions to the data.

\MakeTable{lccccc}{12.5cm}{New double-mode Cepheids in OGLE-II LMC sample}%1
{\hline
\noalign{\smallskip}
 OGLE ID          & $P_0$\thinspace [day]
                             & $P_1$\thinspace [day]
                                        & $P_1/P_0$
                                                  & $A_1/A_0$
                                                           & Type \\
\noalign{\smallskip}
\hline
\noalign{\smallskip}
 SC2--263415      & 3.420377 & 2.455273 & 0.71784 &  27.11 & FU/FO \\
 SC11--250925     & 7.864120 & 5.565078 & 0.70765 &  11.11 & FU/FO \\
 SC17--126402$^a$ & 3.209709 & 2.294942 & 0.71500 &  24.31 & FU/FO \\
 SC20--100652     & 3.747624 & 2.715290 & 0.72454 &  16.65 & FU/FO \\
\noalign{\smallskip}
\hline
\noalign{\smallskip}
                  & $P_1$\thinspace [day]
                             & $P_2$\thinspace [day]
                                        & $P_2/P_1$
                                                  & $A_2/A_1$
                                                           &       \\
\noalign{\smallskip}
\hline
\noalign{\smallskip}
 SC20--112788$^b$ & 0.737740 & 0.594287 & 0.80555 & ~~0.09 & FO/SO \\
\noalign{\smallskip}
\hline
\noalign{\smallskip}
\multicolumn{6}{p{9cm}}{$^a$ Additional mode at P=1.429530\thinspace day detected (see Section~6).}\\
\multicolumn{6}{p{9cm}}{$^b$ Both radial modes modulated (see Section~7).}\\
}

In Fig.\thinspace 1 and Fig.\thinspace 2 we present the period
ratio {\it vs.} period diagrams (so called Petersen diagrams) for
all double-mode Cepheids detected in the LMC (Alcock {\it
et~al.}\thinspace 1995, 1999, 2003; Soszy\'nski {\it
et~al.}\thinspace 2000). For comparison, we also display the FU/FO
Cepheids observed in the Small Magellanic Cloud (Alcock {\it
et~al.}\thinspace 1997; Udalski {\it et~al.}\thinspace 1999a) and
in the Galaxy (Pardo \& Poretti 1997; Antipin 1997a, 1998;
Berdnikov \& Turner 1998; Wils \& Otero 2004)\footnote{For DZ~CMa
pulsation periods have been redetermined with new data of
Berdnikov (2008).}. Because of different metallicities, the
$P_1/P_0$ period ratios in these three stellar systems are
somewhat different. Four of the newly identified double-mode
Cepheids fit to their respective Petersen diagrams very well. For
the fifth object, SC20--100652, the period ratio $P_1/P_0$ is
rather high and is close to the value expected for an SMC star.
This suggests that SC20--100652 has a lower metallicity than other
LMC Cepheids. We also note that, the new FU/FO double-mode
pulsator SC11--250925 has the longest periods among currently
known variables of this class.

\section{Triple-mode Cepheids}%4

In two of the previously known FO/SO double-mode Cepheids a third
strong periodicity was found. The period ratio of $P_3/P_2 =
0.840$, the same in both Ce\-phe\-ids, identifies the new mode as
a third radial overtone. In Fig.\thinspace 3 we present the
prewhitening sequence for one of these stars. The third overtone
is detected very securely, in fact it is stronger than the second
overtone.

\MakeTable{lccccccc}{12.5cm}{Triple-mode Cepheids in OGLE-II LMC sample}%2
{\hline
\noalign{\smallskip}
OGLE ID      & $P_1$\thinspace [day]
                        & $P_2$\thinspace [day]
                                   & $P_3$\thinspace [day]
                                              & $P_2/P_1$
                                                        & $P_3/P_2$
                                                                  & $A_2/A_1$
                                                                          & $A_3/A_1$ \\
\noalign{\smallskip}
\hline
\noalign{\smallskip}
 SC3--360128 & 0.541279 & 0.436046 & 0.366300 & 0.80559 & 0.84005 & 0.167 & 0.311 \\
 SC5--338399 & 0.579510 & 0.466624 & 0.392116 & 0.80520 & 0.84033 & 0.111 & 0.178 \\
\noalign{\smallskip}
\hline
}

Pulsation properties of the two triple-mode Cepheids are listed in
Table~2. In Fig.\thinspace 4 we display both variables in the
Cepheid $P-L$ plot constructed for extinction insensitive
Wessenheit index, $W_I = I-1.55\times (V-I)$ (Madore \& Freedman
1991). The stars fit the $P-L$ relation very well, proving that
apart from multiperiodic pulsations, they are both normal
Pop.\thinspace I Cepheids belonging to the LMC.

Discovery of the triple-mode pulsators listed in Table~2 was
originally announced by Moskalik, Ko{\l}aczkowski \& Mizerski
(2004). Three more triple-mode Cepheids have been identified very
recently by Soszy\'nski {\it et~al.}\thinspace (2008a), bringing
the number of objects in the class to five. This new type of
multimode Cepheid pulsations, although extremely rare, is very
important for testing stellar models. With three radial modes
observed and their periods accurately measured, seismological
analysis yields tight constraints on Cepheid parameters (mass,
luminosity, metalli\-city). This imposes constraints on Cepheid
evolutionary tracks, including interesting limits on convective
overshooting from the core (Moskalik \& Dziembowski 2005).

\section{Analysis of FO and FU Cepheids}%5

\subsection{Nonradial modes in first overtone Cepheids}%5.1

The OGLE-II catalog of LMC Cepheids (Udalski {\it et~al.}\thinspace
1999b) lists 462 first overtone variables. We detected resolved
secondary frequencies in 42 of them. This constitutes 9\% of the
sample. We consider two frequencies to be resolved, if they differ
by more than $\Delta f = 2/T$, where $T$ is the length of the data.
In case of OGLE-II photometry, this means $\Delta f > 0.0017 -
0.0020$\thinspace c/d, depending on the star. Our criterion is more
conservative than usually adopted in other studies (i.e. Alcock {\it
et~al.}\thinspace 2003; Nagy \& Kov\'acs 2006).

The complete inventory of resolved FO Cepheids is presented in
Table~3. Following notation originally introduced for the RR~Lyr
variables (Alcock {\it et~al.}\thinspace 2000), we call these stars
FO-$\nu$ Cepheids. Consecutive columns of Table~3 give OGLE number
of the star, primary and secondary periods $P_1$ and $P_{\nu}$,
frequency difference $\Delta f = f_{\nu} - f_1$, period ratio
$P_{\nu}/P_1$ and amplitude ratio $A_{\nu}/A_1$.

\MakeTable{lcccccl}{12.5cm}{FO-${\nu}$ Cepheids in OGLE-II LMC sample}%3
{\hline
\noalign{\smallskip}
OGLE ID       & $P_1$\thinspace [day]
                         & $P_{\nu}$\thinspace [day]
                                    & $\Delta f$\thinspace [c/d]
                                               & $P_{\nu}/P_1$
                                                         & $A_{\nu}/A_1$
                                                                 & Remarks \\
\noalign{\smallskip}
\hline
\noalign{\smallskip}
 SC1--201683  & 2.319211 & 2.340033 & -0.00384 & 1.00898 & 0.048 &         \\
              &          & 3.069683 & -0.10542 & 1.32359 & 0.022 &         \\
 SC1--229541  & 1.216360 & 1.359217 & -0.08641 & 1.11745 & 0.043 & a       \\
 SC2--158672  & 2.313334 & 1.477135 & ~0.24471 & 0.63853 & 0.032 & b       \\
 SC2--208897  & 2.417847 & 2.346886 & ~0.01251 & 0.97065 & 0.074 &         \\
 SC2--283723  & 1.308907 & 1.425531 & -0.06250 & 1.08910 & 0.070 &         \\
 SC3--274410  & 2.446161 & 2.981644 & -0.07342 & 1.21891 & 0.043 &         \\
              &          & 3.099460 & -0.08617 & 1.26707 & 0.033 & a       \\
 SC3--421512  & 3.186057 & 4.652311 & -0.09892 & 1.46020 & 0.055 &         \\
 SC4--36200   & 3.476962 & 3.912773 & -0.03203 & 1.12534 & 0.054 &         \\
              &          & 4.057330 & -0.04114 & 1.16692 & 0.038 &         \\
 SC4--131738  & 3.122245 & 3.780949 & -0.05580 & 1.21097 & 0.032 &         \\
 SC4--295932  & 4.166232 & 2.633729 & ~0.13967 & 0.63216 & 0.023 &         \\
 SC5--75989   & 2.238255 & 2.949657 & -0.10775 & 1.31784 & 0.091 & s1      \\
 SC5--138031  & 2.681369 & 3.249411 & -0.06520 & 1.21185 & 0.029 &         \\
 SC6--135695  & 1.864922 & 1.949614 & -0.02329 & 1.04541 & 0.037 &         \\
 SC6--135716  & 2.838799 & 2.780248 & ~0.00742 & 0.97938 & 0.064 &         \\
 SC6--267289  & 1.845746 & 1.906899 & -0.01738 & 1.03313 & 0.039 &         \\
 SC6--363194  & 2.797085 & 3.727012 & -0.08920 & 1.33246 & 0.043 & a       \\
 SC7--344559  & 2.062992 & 2.140660 & -0.01759 & 1.03765 & 0.068 &         \\
 SC8--205108  & 3.515988 & 3.905659 & -0.02884 & 1.11083 & 0.026 &         \\
 SC8--224964  & 2.866473 & 3.189296 & -0.03531 & 1.11262 & 0.041 & a       \\
 SC9--216934  & 4.001017 & 4.997407 & -0.04983 & 1.24903 & 0.031 &         \\
 SC9--230584  & 4.395271 & 5.672324 & -0.05122 & 1.29055 & 0.059 & c       \\
              &          & 5.860579 & -0.05689 & 1.33338 & 0.028 & a       \\
 SC10--95827  & 1.974513 & 2.070361 & -0.02345 & 1.04854 & 0.076 &         \\
 SC10--132645 & 1.576676 & 1.594462 & -0.00708 & 1.01128 & 0.053 &         \\
 SC10--259946 & 5.075153 & 6.183343 & -0.03531 & 1.21836 & 0.039 &         \\
 SC13--165223 & 2.043172 & 2.039852 & ~0.00080 & 0.99838 & 0.040 & d       \\
              &          & 1.230847 & ~0.32301 & 0.60242 & 0.041 &         \\
 SC13--242700 & 3.452472 & 3.262793 & ~0.01684 & 0.94506 & 0.060 & a       \\
              &          & 3.363016 & ~0.00771 & 0.97409 & 0.077 &         \\
 SC14--46315  & 2.315774 & 1.442102 & ~0.26161 & 0.62273 & 0.023 & b       \\
 SC15--170744 & 4.992984 & 3.148984 & ~0.11728 & 0.63068 & 0.055 &         \\
 SC16--37119  & 2.795951 & 3.520911 & -0.07364 & 1.25929 & 0.051 &         \\
 SC16--177823 & 1.918044 & 1.892260 & ~0.00710 & 0.98656 & 0.046 &         \\
 SC16--194279 & 2.065767 & 2.146254 & -0.01815 & 1.03896 & 0.059 &         \\
 SC16--230207 & 3.418163 & 4.416698 & -0.06614 & 1.29213 & 0.039 & s2      \\
 SC17--39484  & 3.541888 & 4.063954 & -0.03627 & 1.14740 & 0.055 &         \\
              &          & 4.104648 & -0.03871 & 1.15889 & 0.045 &         \\
 SC17--39517  & 2.113535 & 2.149961 & -0.00802 & 1.01725 & 0.065 &         \\
              &          & 1.326966 & ~0.28046 & 0.62784 & 0.046 &         \\
\noalign{\smallskip}
\hline
}

\setcounter{table}{2}%3-continued
\MakeTable{lcccccl}{12.5cm}{Concluded}
{\hline
\noalign{\smallskip}
OGLE ID       & $P_1$\thinspace [day]
                         & $P_{\nu}$\thinspace [day]
                                    & $\Delta f$\thinspace [c/d]
                                               & $P_{\nu}/P_1$
                                                         & $A_{\nu}/A_1$
                                                                 & Remarks \\
\noalign{\smallskip}
\hline
\noalign{\smallskip}
 SC17--80220  & 2.258854 & 2.829748 & -0.08931 & 1.25274 & 0.058 &         \\
 SC17--146711 & 1.883131 & 1.137820 & ~0.34784 & 0.60422 & 0.049 &         \\
 SC17--171481 & 2.537651 & 3.707561 & -0.12435 & 1.46102 & 0.066 &         \\
 SC17--211310 & 2.023579 & 2.087774 & -0.01519 & 1.03172 & 0.054 & s3      \\
              &          & 2.593589 & -0.10861 & 1.28168 & 0.037 & a       \\
 SC18--144653 & 4.158193 & 5.222752 & -0.04902 & 1.25602 & 0.049 &         \\
 SC18--208875 & 3.928946 & 3.508220 & ~0.03052 & 0.89292 & 0.202 &         \\
 SC19--74265  & 2.364944 & 2.790232 & -0.06445 & 1.17983 & 0.046 & a       \\
              &          & 3.080875 & -0.09826 & 1.30273 & 0.060 &         \\
 SC20--83423  & 2.068067 & 2.108720 & -0.00932 & 1.01966 & 0.053 &         \\
\noalign{\smallskip}
\hline
\multicolumn{7}{p{10.3cm}}{Remarks: \ a) Marginal detection; \ b) Unresolved
                           residual power at primary frequency; \ c) Unresolved
                           residual power at secondary frequency; \ d) Analyzed
                           with MACHO data; \ s1) Same as SC6--422324; \
                           s2) Same as SC17--33268; \ s3) Same as SC18--29237.}
}

In most of the FO-$\nu$ Cepheids only one secondary frequency was
detected, but in several variables two frequencies were found. In
all cases they have extremely low amplitudes. With the exception of
a singe star (SC18--208875), the amplitude ratio $A_{\nu}/A_1$ is
always below 0.1, with the average value of 0.048. We note, that
secondary peaks detected in LMC first overtone RR~Lyr stars are
several times higher, with $A_{\nu}/A_1 = 0.31$ on average (Nagy \&
Kov\'acs 2006).

The secondary frequencies in FO-$\nu$ Cepheids come in two
different flavours. In 37 variables they are located close to the
primary (radial) frequency $f_1$, within $|\Delta f| <
0.13$\thinspace c/d. Representative examples of such behaviour are
displayed in Fig.\thinspace 5. In 84\% of cases, secondary
frequencies are lower than the primary one. When two secondary
peaks are detected, they always appear on the same side of the
primary peak. The observed frequency pattern cannot be explained
by amplitude or phase modulation of the radial mode. A periodic
modulation always produces an equally spaced frequency multiplet
(triplet, quintuplet, etc.), centered on the primary peak. It is
also easy to check, that measured period ratios are incompatible
with those of the radial modes. This implies, that secondary
frequencies detected close to the radial mode of the FO Cepheids
must correspond to {\it nonradial modes of oscillations}. This is
the first detection of nonradial modes in this type of stars.

In seven FO-$\nu$ Cepheids secondary periodicities of a different
type are found: we see a {\it high frequency mode}, with
characteristic period ratio of $P_{\nu}/P_1 = 0.60 - 0.64$.
Examples of this behaviour are displayed in Fig.\thinspace 6. The
two types of se\-condary modes are not mutually exclusive. In two
objects (SC13--165223 and SC17--39517) both a high frequency
secondary peak and a secondary peak close to the primary frequency
are present.

The mysterious period ratio of $0.60-0.64$ cannot be explained in
terms of radial modes, neither. Observationally determined period
ratio of the third and first radial overtones is $P_3/P_1 = 0.6766
- 0.6773$ (see Table~2; Soszy\'nski {\it et~al.}\thinspace 2008a).
The value of $P_4/P_1$ can be estimated theoretically. Assuming
validity of $p$-mode asymptotic relation $f_4 - f_3 = f_3 - f_2$,
we get $P_4/P_3 = (2 - P_3/P_2)^{-1}$. With $P_3/P_2 = 0.840$ (see
Table~2) we find $P_4/P_3 = 0.862$, thus $P_4/P_1 = 0.5835$. This
estimate agrees nicely with $P_4/P_1 = 0.5898$ derived from
tentative detection of the fourth overtone in one of the LMC
Cepheids (Soszy\'nski {\it et~al.}\thinspace 2008a). With
$P_{\nu}/P_1 = 0.60 - 0.64$, the high frequency secondary peaks in
FO-$\nu$ Cepheids cannot correspond to neither the third nor the
fourth radial overtone. They have to be attributed to nonradial
modes. These modes appear at frequencies $0.02-0.06$\thinspace c/d
below frequency of the (unobserved) fourth radial overtone.
Interestingly, this frequency difference is comparable to $\Delta
f$ for the nonradial modes detected in the vicinity of the first
overtone.

In Fig.\thinspace 7 we show the distribution of frequency
differences, $\Delta f$, for the LMC FO-$\nu$ Cepheids and, for
comparison, for the LMC first overtone Blazhko RR~Lyr stars
(RRc-BL stars). The two distributions are very similar. FO-$\nu$
Cepheids show stronger preference for negative $\Delta f$, but
otherwise, in both types of overtone pulsators nonradial modes are
found in similar distances from the radial mode. The only
difference between the two histograms is presence of high
frequency se\-condary modes ($P_{\nu}/P_1 = 0.60 - 0.64$), which
were detected in FO Cepheids, but not in RRc stars.

\subsection{First overtone Cepheids with unresolved residual power}%5.2

In 23 first overtone Cepheids we found after prewhitening a
significant resi\-dual power unresolved from primary pulsation
frequency. This is a signature of a slow amplitude and/or phase
variability, occuring on a timescale comparable or longer than the
$\sim 1100$\thinspace days length of OGLE-II data. We reanalyzed
all these stars with the MACHO $b$ photometry (Allsman \& Axelrod
2001), which offered a much longer timebase of 2700-2800\thinspace
days. MACHO data was cleaned before analysis. We removed all
datapoints with formal errors more than 5 times larger than the
average formal error in the dataset. We also rejected all
measurements differing from the mean brightness by more than 5
standard deviations (of the unphased data). After cleaning, the
data was analyzed in a standard way. Remnant power was resolved
with MACHO photometry only in one additional Cepheid,
SC13--165223. This object is included in Table~3. Remaining 22
Cepheids with unresolved residual power are listed in Table~4.
Following notation introduced by Alcock {\it et~al.}\thinspace
(2000), we call these variables FO-PC Cepheids.

\MakeTable{lccl}{12.5cm}{FO-PC Cepheids in OGLE-II LMC sample}%4
{\hline
\noalign{\smallskip}
OGLE ID       & $P_1$\thinspace [day]
                         & $A_1$\thinspace [mag]
                                 & Remarks \\
\noalign{\smallskip}
\hline
\noalign{\smallskip}
 SC1--158032  & 2.161982 &  ---  & ab,c    \\
 SC1--324986  & 2.209283 & 0.173 & ab,s1   \\
 SC2--248540  & 1.823789 &  ---  & a,d     \\
 SC4--152290  & 1.887862 & 0.082 & ab      \\
 SC8--76179   & 2.462125 & 0.183 & b,s2    \\
 SC9--153398  & 0.947292 & 0.087 & ab      \\
 SC10--269402 & 1.912859 & 0.114 & ab      \\
 SC11--338308 & 1.743443 & 0.053 & ab      \\
 SC13--74204  & 1.369660 & 0.079 & a       \\
 SC14--148402 & 2.193230 & 0.209 & b       \\
 SC15--31558  & 1.554964 & 0.046 & a       \\
 SC15--158999 & 1.100853 & 0.266 & b       \\
 SC16--99240  & 2.437482 & 0.152 & b       \\
 SC17--15475  & 2.012359 & 0.200 & ab      \\
 SC17--117748 & 3.572147 & 0.160 & b       \\
 SC17--122399 & 2.783800 & 0.067 & a       \\
 SC17--214859 & 1.954504 & 0.206 & b,s3    \\
 SC18--141019 & 2.270553 & 0.192 & b       \\
 SC18--202349 & 2.432908 & 0.179 & b,s4    \\
 SC21--85282  & 3.372482 & 0.167 & ab      \\
 SC21--106503 & 2.535978 & 0.184 & ab      \\
 SC21--136158 & 3.774050 & 0.153 & ab      \\
\noalign{\smallskip}
\hline
\multicolumn{4}{p{6.2cm}}{Remarks: \ a) Amplitude variability; \ b) Phase
                          variability; \ c) no residual power detected in MACHO
                          data; \ d) no MACHO data available; \ s1) Same as
                          SC16--57446; \ s2) Same as SC9--372261; \ s3) Same as
                          SC18--33591; \ s4) Same as SC19--48662.}
}

For closer examination of slow amplitude/phase variability in
FO-PC Cepheids we divided the data into 10 subsets, each spanning
10\% of the total timebase, and then fitted Eq.(1) to each subset
separately. This way we were able to follow the time evolution of
amplitude and phase of the radial mode. In 8 stars only parabolic
trends in pulsation phases were found. These Cepheids most likely
undergo a se\-cular period change. In remaining 14 FO-PC Cepheids
variability of amplitudes is clearly seen, in most cases
associated with variability of phases. In Fig.\thinspace 8 we show
typical examples of both behaviours. With currently available data
we cannot make any firm statement about the cause of the observed
long term trends. We only note, that in {\it all} FO-PC Cepheids
the observed changes of phases and amplitudes are orders of
magnitude too fast to be explained by stellar evolution. We
recall, that fast phase variations of nonevolutionary origin are
not unique to Cepheids of the LMC, they are also observed in many
FO Cepheids in the Galaxy (Berdnikov {\it et~al.}\thinspace 1997,
their Fig.\thinspace 3). On the other hand, secular amplitude
variations are rare in Galactic FO Cepheids, with Polaris being
the only well documented case ({\it e.g.}\thinspace Lee {\it
et~al.}\thinspace 2008; Bruntt {\it et~al.}\thinspace 2008).

\subsection{Fundamental mode Cepheids}%5.3

The OGLE-II catalog of LMC Cepheids (Udalski {\it et~al.}\thinspace
1999b) list 718 fundamental mode pulsators. We searched all of them
for secondary periodicities. We found {\it no nonradial modes} in
the FU Cepheids of the LMC. This result is highly significant,
considering that the FU Cepheid sample is much larger than the FO
Cepheid sample. Long-term amplitude or phase variability is also
extremely rare in the FU Cepheids. We found this behaviour only in
one star (SC17--94969).

\subsection{Incidence rate of nonradial modes}%5.4

In Table~5 we present the inventory of different types of pulsators
identified in our survey of FO and FU Cepheids in the LMC. For each
type of variability we give the number of stars found, $N_i$, and
the estimated incidence rate in the population with its statistical
error. The errors are calculated from the assumption of Poisson
distribution ({\it cf.} Alcock {\it et~al.}\thinspace 2003):

\begin{equation}
\sigma_{N_i/N}=  {1\over N} \sqrt{{N_i(N-N_i)}\over N},
\end{equation}

\medskip

\noindent where $N$ is the total number of FO or FU Cepheids.

\MakeTable{llrr}{12.5cm}{Variability types in FO and FU Cepheids of OGLE-II LMC sample}%5
{\hline
\noalign{\smallskip}
Type      & Description                         & Number
                                                         & Inc. rate~ \,    \\
\noalign{\smallskip}
\hline
\noalign{\smallskip}
 FO-S     & Single-periodic overtone Cepheid    & 397~~~ & $85.9 \pm 1.6$\% \\
 FO-$\nu$ & FO with nonradial modes             &  42~~~ &  $9.1 \pm 1.3$\% \\
 FO-PC    & FO with variable phase/amplitude    &  22~~~ &  $4.8 \pm 1.0$\% \\
 FO-Misc  & FO with some miscellany             &   1~~~ &  $0.2 \pm 0.2$\% \\
\noalign{\smallskip}
\hline
\noalign{\smallskip}
 FU-S     & Single-periodic fundamental Cepheid & 716~~~ & $99.6 \pm 0.2$\% \\
 FU-$\nu$ & FU with nonradial modes             &  --~~~ &  --------~~~~ \\
 FU-PC    & FU with variable phase/amplitude    &   1~~~ &  $0.1 \pm 0.1$\% \\
 FU-Misc  & FU with some miscellany             &   1~~~ &  $0.1 \pm 0.1$\% \\
\noalign{\smallskip}
\hline
}

The most significant result of our survey is the detection of
nonradial modes in LMC Cepheids, but only in those which pulsate
in the first overtone. In Fig.\thinspace 9 we display all Cepheids
with nonradial modes on the Wessenheit index $P-L$ dia\-gram. All
variables classified as FO-${\nu}$ Cepheids are indeed firmly
located on the first overtone sequence. Their distribution with
the pulsation period is not uniform, though. FO-${\nu}$ pulsators
are not found for $ P_1 < 1.2$\thinspace day. This suggests that
the incidence rate of nonradial modes might depend on the period.
To test this hypo\-thesis we divided the FO Cepheid population
into several period bins and estimated the incidence rates in each
bin separately. Results are displayed in Fig.\thinspace 10. The
incidence rate of FO-${\nu}$ pulsators systematically increases
with period, reaching $19 \pm 5$\% for $\log P_1 > 0.5$. We think,
that this is most likely a selection effect. The nonradial modes,
which have very small amplitudes, are progressively more difficult
to detect in Cepheids with shorter periods {\it i.e.}\thinspace
with lower luminosities. This interpretation implies, that the
true incidence rate of FO-${\nu}$ Cepheids might be significantly
higher than 9\% given in Table~5.

\subsection{Comparison with RR~Lyrae stars}%5.5

In the last decade nonradial modes were detected in many RR~Lyr
stars belonging to various stellar systems. Since pulsations of
Cepheids and of RR~Lyr stars are in many ways similar, it is
interesting to compare the properties of nonradial modes in these
two types of variables. One similarity has already been discussed
in Section~5.1: in both types of pulsators, nonradial modes appear
in essentially the same frequency distances from the radial mode
(Fig.\thinspace 7). However, there are also striking differences
between the two groups of stars:

\begin{itemize}
\item Amplitudes of nonradial modes in classical Cepheids are very low, on
      average almost an order of magnitude {\it lower} than in RR~Lyr stars.
\item In RR~Lyr variables nonradial modes are detected both in fundamental
      (RRab) and in first overtone pulsators (RRc). In fact, in all
      studied stellar systems they are found in the RRab stars
      either equally frequently (Soszy\'nski {\it et~al.}\thinspace
      2002) or much more frequently (Soszy\'nski {\it
      et~al.}\thinspace 2003; Mizerski 2003). In sharp contrast, in
      classical Cepheids nonradial modes are detected {\it only in
      the first overtone pulsators}. The difference between
      fundamental mode Cepheids and RR~Lyr stars is even more
      striking, considering that for overtone pulsators incidence
      rates of nonradial modes are roughly the same: 9.1\% in FO
      Cepheids and $9.6-12.1$\% in RRc stars (Nagy \& Kov\'acs 2006,
      Mi\-zer\-ski 2003).
\item When two secondary frequencies are found in an RR~Lyr star, they
      usually form together with the primary frequency an equally spaced
      triplet, centered on the primary (radial) peak. Such equidistant triplets
      are {\it never} observed in Cepheids.
\item In RR~Lyr stars nonradial modes are detected only in close vicinity of
      the primary pulsation mode. This is also the case for most of the
      FO-$\nu$ Cepheids, but in several variables a high frequency mode with
      $P_{\nu}/P_1 = 0.60-0.64$ is found. Such high frequency modes were
      {\it not detected} in RR~Lyr pulsators.
\end{itemize}

\subsection{Can secondary frequencies result from blending ?}%5.6

In Section~2 we describe how to identify spurious secondary
frequencies, which have appeared through a contamination of DIA
photometry with light of neighbouring variables. This procedure
cannot identify cases of real blending, resulting from the angular
resolution limit of the observations. This raises a legitimite
question, whether secondary periodicities discussed in this paper
originate in Cepheids or whether they are introduced by blending
of Cepheids with other variable stars. Although the latter
possibility cannot be ruled out in any individual case, we are
convinced that majority of nonradial modes detected in FO-$\nu$
Cepheids cannot be explained by blending. Our arguments are
statistical. First, observed distribution of frequency
differences, $\Delta f = f_{\nu} - f_1$, is not random
(Fig.\thinspace 7). There is no reason why blending should
introduce frequencies only in the vicinity of a radial mode or at
any particular narrow range of period ratios. Second, we do not
find nonradial modes in Cepheids, which pulsate in the fundamental
mode. There is no reason why blending should affect stars
pulsating in one mode, but not in the other mode. Third, blending
should be more substantial ({\it i.e.}\thinspace relative flux
contribution should be larger) in case of fainter Cepheids
(Mochejska 2002). Therefore, secondary pe\-rio\-dicities
introduced by blending should be found preferentially in short
period pulsators (where their amplitudes should be higher). The
statistics of FO-$\nu$ Cepheids shows just the opposite trend
(Fig.\thinspace 10). Finally, probability of blending should
increase proportionally to the density of stars in the field (Kiss
\& Bedding 2005). This is not the case for the FO-$\nu$ Cepheids.
In 8 densest LMC fields (SC2-SC9; 189 FO Cepheids in total) and in
8 least dense LMC fields (SC12, SC14, SC15, SC17-SC21; 157 FO
Cepheids in total) the incidence rates of FO-$\nu$ pulsations are
$10.6\pm 2.2$\% and $8.9\pm 2.3$\%, respectively. The two values
are statistically the same within $0.6\sigma$, despite two-fold
difference in average number of stars per field. From all these
evidences we conclude that nonradial modes detected in the
FO-$\nu$ Cepheids do not result from blending, but are intrinsic
to Cepheids themselves.

\section{Nonradial modes in FU/FO double-mode Cepheids}%6

OGLE-II catalog of double-mode LMC Cepheids (Soszy\'nski {\it
et~al.}\thinspace 2000) lists 19 FU/FO pulsators. Discovery of
four new members (Section~3) brings the total number of Cepheids
in this class to 23. We found nonradial modes in three of them.
These are the first detections of nonradial modes in the FU/FO
double-mode Cepheids. In the following, we will call these
variables FU/FO-$\nu$ Cepheids. The stars are listed in Table~6.
In the fifth column of the Table we give the frequency difference
between the nonradial mode and the first radial overtone, $\Delta
f = f_{\nu} - f_1$. The prewhitened power spectra of the
FU/FO-$\nu$ variables are displayed in Fig.\thinspace 11. In the
first two Cepheids, the secondary mode appears very close to the
first overtone radial mode. The values of $\Delta f$ are very
similar to those observed in the FO-$\nu$ Cepheids. In the third
star, the secondary mode is found at high frequency, with
$P_{\nu}/P_1 = 0.6229$. This is the same puzzling period ratio,
which has been found in several FO-$\nu$ pulsators. Clearly,
nonradial modes excited in the FU/FO-$\nu$ Cepheids are somehow
related to the first radial overtone and their frequencies seem to
be drawn from the same distribution as in the case of the FO-$\nu$
Cepheids.

\MakeTable{lcccccc}{12.5cm}{FU/FO-${\nu}$ double-mode Cepheids in OGLE-II LMC sample}%6
{\hline
\noalign{\smallskip}
OGLE ID       & $P_0$\thinspace [day]
                         & $P_1$\thinspace [day]
                                    & $P_{\nu}$\thinspace [day]
                                               & $\Delta f$\thinspace [c/d]
                                                          & $P_{\nu}/P_1$
                                                                    & $A_{\nu}/A_1$ \\
\noalign{\smallskip}
\hline
\noalign{\smallskip}
 SC6--6       & 4.841143 & 3.453214 & 3.581206 & -0.01035 & 1.03706 & 0.057 \\
 SC15--69667  & 1.904133 & 1.376951 & 1.517867 & -0.06742 & 1.10234 & 0.052 \\
 SC17--126402 & 3.209709 & 2.294942 & 1.429530 & ~0.26379 & 0.62290 & 0.040 \\
\noalign{\smallskip}
\hline
}

\section{Blazhko effect in FO/SO double-mode Cepheids}%7

Including one new variable discovered in the current paper
(Section~3), there are 56 FO/SO double-mode Cepheids in the
OGLE-II catalog (Soszy\'nski {\it et~al.}\thinspace 2000). In
one-third of these stars we detected residual signal close to one
or both primary (radial) pulsation frequencies. In all cases,
however, the secondary frequencies were not resolved from the
primary ones. Therefore, the frequency analysis of all FO/SO
pulsators was repeated with MACHO $b$ photometry (Allsman \&
Axelrod 2001), which provided $\sim$2.5 times better frequency
resolution than OGLE-II. The MACHO data was cleaned in a manner
described in Section~5.2 and then analyzed in a standard way. At
this stage the OGLE-II sample was supplemented by 51 additional
LMC FO/SO double-mode Cepheids discovered by MACHO outside the
area covered by the OGLE-II survey (Alcock {\it et~al.}\thinspace
1999, 2003). Our final FO/SO Cepheid sample consisted of 107
objects.

With the MACHO data, residual power close to the primary pulsation
frequencies was detected in 37 FO/SO double-mode Cepheids (35\% of
the sample). In 20 of these stars, which constitute 19\% of the
sample, we were able to resolve this power into individual
frequencies. Two frequencies are considered resolved if they differ
by $\Delta f > 0.000715$\thinspace c/d ({\it i.e.}\thinspace
$1/\Delta f < 1400$\thinspace days). All resolved FO/SO Cepheids are
listed in Table~7.

\subsection{Frequency domain}%7.1

Resolved FO/SO double-mode Cepheids display a very characteristic
fre\-quen\-cy pattern. In most cases, in the vicinity of a radial
frequency we detect two secondary peaks. They are located on the
opposite sides of the primary (radial) peak and together with the
primary peak they form an {\it equally spaced frequency triplet}.
Such a structure is usually found around {\it both} radial modes.
An example of this pattern is shown in Fig.\thinspace 12. In
several cases instead of triplets we find only doublets. This
happens in objects, in which secondary frequencies are detected
with the lowest signal-to-noise ratio. Therefore, it is most
likely that these doublets are in fact parts of triplets, with the
missing third component hidden in the noise. When signal-to-noise
ratio is high we encounter an opposite situation. In two FO/SO
Cepheids we find {\it equally spaced frequency quintuplets}
centered on the second overtone. Consecutive prewhitening reveals
incomplete quintuplets in six other stars, including three
variables where they apper around the first overtone. In
Fig.\thinspace 13 we display representative examples of triplets
and quintuplets observed in the FO/SO pulsators.

The secondary peaks are always very small, with amplitudes of
15\thinspace mmag on average and exceeding 40\thinspace mmag only in
one star. The side peaks of the multiplets are usually not equal.
For the first overtone, the low frequency components of
triplets/quintuplets are always higher than the high frequency ones.
This is also true for 74\% of multiplets centered on the second
overtone.

The separations of components in the two triplets/quintuplets,
$\Delta f_1$ and $\Delta f_2$, are also very small and never exceed
0.00143\thinspace c/d. This corresponds to the beat (or modulation)
period of $P_B = 1/\Delta f > 700$\thinspace day. The lower limit
for $\Delta f$ (upper limit for $P_B$) is currently given by
resolution of the data. The physical upper limit for the beat
(modulation) period $P_B$ is unknown.

When present around both radial modes, the two multiplets have
nearly the same frequency spacings. Specifically, the difference
between the two spacings, $|\Delta f_2\! -\!\Delta f_1|$ is always
below $8\times 10^{-5}$\thinspace c/d. According to Monte Carlo
simulations of Alcock {\it et~al.}\thinspace (2000, 2003), for
MACHO data such a difference is well within deviations expected
from observational noise. In other words, triplets/quintuplets
around both radial modes of a FO/SO Cepheid have {\it the same
frequency separation} within accuracy of the data.

Frequency separation $\Delta f$ can take any value below
0.00143\thinspace c/d, but some va\-lues are more likely than
others. This is clearly visible in Fig.\thinspace 14. Although
statistics is rather small, the distribution of $\Delta f$ shows a
bimodal shape with pronounced maxima at $\sim 0.00095$\thinspace
c/d and $\sim 0.00125$\thinspace c/d. These correspond to
preferred beat (modulation) periods of $P_B\sim 1050$\thinspace
days and $P_B\sim 800$\thinspace days.

\begin{landscape}

\MakeTable{lcccccccccccl}{12.5cm}{Blazhko FO/SO double-mode Cepheids in combined OGLE-II + MACHO LMC sample}%7
{\hline
\noalign{\smallskip}
             &          &          &         &         &       &       & $A_1^{-}$
                                                                               & $A_1^{- -}$
                                                                                       &       & $A_2^{-}$
                                                                                                       & $A_2^{- -}$
                                                                                                               &         \\
\noalign{\smallskip}
OGLE ID      & $P_1$    & $P_2$    & $\Delta f_1$
                                             & $\Delta f_2$
                                                       & $P_B$ & $A_1$ & $A_1^{+}$
                                                                               & $A_1^{++}$
                                                                                       & $A_2$ & $A_2^{+}$
                                                                                                       & $A_2^{++}$
                                                                                                               & Remarks \\
\noalign{\smallskip}
             & [day]    & [day]    &  [c/d]  &  [c/d]  & [day] & [mag] & [mag] & [mag] & [mag] & [mag] & [mag] &         \\
\noalign{\smallskip}
\hline
\noalign{\smallskip}
SC1--44845   & 0.951981 & 0.766013 & 0.00127 & 0.00125 & ~~794 & 0.165 & 0.021 &  ---  & 0.046 & 0.021 &  ---  &         \\
             &          &          &         &         &       &       & 0.011 &  ---  &       & 0.010 &  ---  &         \\

SC1--285275  & 0.856624 & 0.689234 & 0.00112 & 0.00109 & ~~902 & 0.177 & 0.031 &  ---  & 0.040 & 0.022 & 0.008 &         \\
             &          &          &         &         &       &       & 0.014 &  ---  &       & 0.013 &  ---  &         \\

SC1--335559  & 0.749802 & 0.603647 & 0.00121 & 0.00128 & ~~789 & 0.227 & 0.019 &  ---  & 0.062 & 0.023 &  ---  &         \\
             &          &          &         &         &       &       &  ---  &  ---  &       &  ---  &  ---  &         \\

SC2--55596   & 0.932530 & 0.751387 & 0.00130 &  ------ & ~~768 & 0.146 & 0.007 &  ---  & 0.039 &  ---  &  ---  & a       \\
             &          &          &         &         &       &       & 0.007 &  ---  &       &  ---  &  ---  &         \\

SC6--142093  & 0.896289 & 0.722058 & 0.00091 & 0.00091 &  1103 & 0.154 & 0.028 &  ---  & 0.043 & 0.017 & 0.012 &         \\
             &          &          &         &         &       &       & 0.013 &  ---  &       & 0.017 & 0.006 &         \\

SC6--267410  & 0.888530 & 0.716773 &  ------ & 0.00117 & ~~857 & 0.120 &  ---  &  ---  & 0.036 &  ---  &  ---  &         \\
             &          &          &         &         &       &       &  ---  &  ---  &       & 0.033 &  ---  &         \\

SC8--10158   & 0.689991 & 0.555699 & 0.00086 & 0.00094 &  1082 & 0.175 & 0.011 &  ---  & 0.031 & 0.019 &  ---  & c, s1   \\
             &          &          &         &         &       &       &  ---  &  ---  &       & 0.012 &  ---  &         \\

SC11--233290 & 1.217532 & 0.978435 & 0.00099 & 0.00104 & ~~991 & 0.186 & 0.019 &  ---  & 0.043 & 0.017 & 0.006 &         \\
             &          &          &         &         &       &       & 0.010 &  ---  &       & 0.010 &  ---  &         \\

SC15--16385  & 0.990417 & 0.795736 & 0.00084 & 0.00088 &  1185 & 0.258 & 0.017 &  ---  & 0.049 & 0.020 &  ---  &         \\
             &          &          &         &         &       &       & 0.016 &  ---  &       &  ---  &  ---  &         \\

SC20--112788 & 0.737740 & 0.594287 & 0.00072 & 0.00072 &  1390 & 0.165 & 0.062 & 0.019 & 0.015 & 0.016 &  ---  &         \\
             &          &          &         &         &       &       & 0.019 &  ---  &       & 0.013 &  ---  &         \\

SC20--138333 & 0.859788 & 0.692171 & 0.00127 & 0.00126 & ~~793 & 0.189 & 0.013 &  ---  & 0.060 & 0.018 &  ---  &         \\
             &          &          &         &         &       &       & 0.008 &  ---  &       & 0.010 &  ---  &         \\
\noalign{\smallskip}
\hline
}

\setcounter{table}{6}%7-continued
\MakeTable{lcccccccccccl}{12.5cm}{Concluded}
{\hline
\noalign{\smallskip}
             &          &        &           &         &       &       & $A_1^{-}$
                                                                               & $A_1^{- -}$
                                                                                       &       & $A_2^{-}$
                                                                                                       & $A_2^{- -}$
                                                                                                               &         \\
\noalign{\smallskip}
MACHO ID     & $P_1$    & $P_2$    & $\Delta f_1$
                                             & $\Delta f_2$
                                                       & $P_B$ & $A_1$ & $A_1^{+}$
                                                                               & $A_1^{++}$
                                                                                       & $A_2$ & $A_2^{+}$
                                                                                                       & $A_2^{++}$
                                                                                                               & Remarks \\
\noalign{\smallskip}
             & [day]    & [day]    &  [c/d]  &  [c/d]  & [day] & [mag] & [mag] & [mag] & [mag] & [mag] & [mag] &         \\
\noalign{\smallskip}
\hline
\noalign{\smallskip}
~ 2.4909.67  & 1.084120 & 0.869989 & 0.00099 & 0.00098 &  1012 & 0.216 & 0.012 &  ---  & 0.055 & 0.013 &  ---  & a,b     \\
             &          &          &         &         &       &       & 0.010 &  ---  &       & 0.009 &  ---  &         \\

13.5835.55   & 0.898730 & 0.722801 & 0.00094 & 0.00095 &  1061 & 0.244 & 0.037 & 0.010 & 0.040 & 0.026 & 0.012 & a       \\
             &          &          &         &         &       &       & 0.021 &  ---  &       & 0.016 &  ---  &         \\

14.9585.48   & 0.935802 & 0.752812 & 0.00091 & 0.00092 &  1098 & 0.139 & 0.024 &  ---  & 0.036 & 0.025 & 0.012 &         \\
             &          &          &         &         &       &       & 0.008 &  ---  &       & 0.010 &  ---  &         \\

17.2463.49   & 0.762933 & 0.614047 & 0.00094 & 0.00093 &  1072 & 0.235 & 0.018 &  ---  & 0.051 & 0.013 &  ---  & d       \\
             &          &          &         &         &       &       & 0.014 &  ---  &       & 0.015 &  ---  &         \\

18.2239.43   & 1.364186 & 1.093324 & 0.00142 & 0.00142 & ~~706 & 0.230 & 0.035 & 0.010 & 0.063 & 0.017 &  ---  & e       \\
             &          &          &         &         &       &       & 0.014 &  ---  &       & 0.015 &  ---  &         \\

22.5230.61   & 0.633077 & 0.510128 &  ------ & 0.00124 & ~~805 & 0.221 &  ---  &  ---  & 0.045 & 0.016 &  ---  &         \\
             &          &          &         &         &       &       &  ---  &  ---  &       & 0.014 &  ---  &         \\

23.2934.45   & 0.734350 & 0.591775 & 0.00129 & 0.00125 & ~~792 & 0.207 & 0.023 &  ---  & 0.055 & 0.031 & 0.015 & a       \\
             &          &          &         &         &       &       &  ---  &  ---  &       & 0.026 & 0.011 &         \\

23.3184.74   & 0.841151 & 0.677788 & 0.00089 & 0.00093 &  1099 & 0.160 & 0.015 &  ---  & 0.048 & 0.011 &  ---  &         \\
             &          &          &         &         &       &       &  ---  &  ---  &       & 0.013 &  ---  &         \\

80.7080.2618 & 0.715873 & 0.577961 & 0.00109 & 0.00110 & ~~914 & 0.168 & 0.010 &  ---  & 0.055 & 0.006 &  ---  & d       \\
             &          &          &         &         &       &       &  ---  &  ---  &       & 0.009 &  ---  &         \\
\noalign{\smallskip}
\hline
\multicolumn{13}{p{16.6cm}}{Remarks: \ a) Unresolved residual power at $f_1$; \
                            b) Unresolved residual power at $f_2$; \
                            c) $f_1^{\,-}$ marginally detected; \
                            d) $f_2^{\,\pm}$ marginally detected; \
                            e) Additional peak at $\Delta f_2 =
                            -0.00076$\thinspace [c/d]; \ s1) Same as
                            SC9--286128.}
}

\end{landscape}

\subsection{Time domain}%7.2

The observed triplet/quintuplet frequency pattern is a Fourier
representation of a periodic modulation of both radial modes, with
the common period $P_B = 1/\Delta f$. In Fig.\thinspace\ 15 we
display this modulation for one of the FO/SO Cepheids. The plot
has been constructed by dividing the data into 10 subsets, each
covering 10\% of the modulation phases, and then fitting radial
frequencies and their linear combinations (Eq.(2)) to each subset
separately. Fig.\thinspace 15 shows that both {\it amplitudes and
phases} of the radial modes undergo modulation. The amplitude
variability is much stronger for the second than for the first
overtone ($\pm 66$\% {\it vs.} $\!\pm 17$\% change). {\it Minimum
amplitude of one mode coincides with maximum amplitude of the
other mode}. Maximum phase of the first overtone occurs just
before maximum of its amplitude. For the second overtone, maximum
phase coincides with minimum amplitude. We note, that the
amplitude of the second overtone is modulated in a distinctively
nonsinusoidal way. This explains presence of quintuplets in the
frequency spectrum of this mode.

Characteristic pattern of periodic modulation displayed in
Fig.\thinspace 15 is common to all FO/SO Cepheids of Table~7. The
phenomenon is strikingly similar to the Blazhko modulation
observed in the RR~Lyrae stars (see {\it e.g.}\thinspace Kurtz
{\it et~al.}\thinspace 2000). Therefore, variables of Table~7 will
be called Blazhko double-mode Cepheids, or FO/SO-BL Cepheids.

\subsection{FO/SO Cepheids with unresolved residual power}%7.3

In 17 FO/SO double-mode Cepheids (16\% of the sample) we find
after pre\-whi\-tening a significant remnant power unresolved from
the primary pulsation frequencies. This is a signature of slow
changes of amplitudes and/or phase of the modes, not resolved
within the length of the data. We call these stars FO/SO-PC
Cepheids. Their list is given in Table~8.

\MakeTable{lccccl}{12.5cm}{FO/SO-PC double-mode Cepheids in combined OGLE-II + MACHO LMC sample}%8
{\hline
\noalign{\smallskip}
OGLE\thinspace /\thinspace MACHO ID
              & $P_1$\thinspace [day]
                         & $P_2$\thinspace [day]
                                    & $A_1$\thinspace [mag]
                                            & $A_2$\thinspace [mag]
                                                    & Remarks \\
\noalign{\smallskip}
\hline
\noalign{\smallskip}
 SC4--176400  & 1.108872 & 0.895139 & 0.089 & 0.023 & ab      \\
 SC4--220148  & 0.740833 & 0.595133 & 0.222 & 0.063 & ab      \\
 SC7--120511  & 1.251130 & 1.003698 & 0.133 & 0.028 & ab      \\
 SC10--204083 & 0.526261 & 0.422879 & 0.245 & 0.030 & b       \\
 SC16--266808 & 1.352920 & 1.081016 & 0.167 & 0.025 & b       \\
 SC17--186042 & 0.610998 & 0.491213 & 0.179 & 0.039 & ab      \\
 SC20--188572 & 1.015249 & 0.814350 & 0.167 & 0.032 & ab      \\
 SC21--12012  & 1.341463 & 1.074876 & 0.181 & 0.057 & ab      \\
 6.6934.67    & 0.920169 & 0.740010 & 0.181 & 0.050 & ab      \\
 11.9348.78   & 0.737799 & 0.594411 & 0.200 & 0.041 & a, s1   \\
 14.9098.35   & 0.727783 & 0.586544 & 0.208 & 0.032 & ab      \\
 15.10428.60  & 0.651848 & 0.525346 & 0.240 & 0.044 & a       \\
 23.3061.82   & 0.595737 & 0.480997 & 0.166 & 0.038 & ab      \\
 24.2853.69   & 0.708844 & 0.571663 & 0.269 & 0.030 & b       \\
 24.2855.80   & 0.593778 & 0.478320 & 0.178 & 0.020 & b       \\
 47.2127.102  & 0.578454 & 0.466643 & 0.227 & 0.024 & ab      \\
 80.7079.62   & 1.347888 & 1.076366 & 0.177 & 0.042 & b       \\
\noalign{\smallskip}
\hline
\multicolumn{6}{p{10cm}}{Remarks: \ a) Residual unresolved power at $f_1$; \
                         b) Residual unresolved power at $f_2$; \ s1) Same as
                        14.9348.3191.}
}

The amplitude and phase variability of each FO/SO-PC Cepheid was
examined with method described in Section~5.2: we divided the data
into 10 subsets, each spanning 10\% of the total timebase, and
then fitted Eq.(2) separately to each subset. In some stars we
found only changes of pulsation phases, but in 10 variables we
also detected clear changes of both amplitudes. In Fig.\thinspace
16 we show typical example of such behaviour. The pattern of
amplitude and phase variability is very much the same as that
displayed in Fig.\thinspace 15. In particular, the amplitudes of
the two radial modes vary in opposite directions. This similarity
suggests, that most of the FO/SO-PC Cepheids experience the same
type of periodic modulation as Blazhko FO/SO Cepheids listed in
Table~7, but on timescales longer than the current data.

\subsection{Incidence rate of Blazhko effect in FO/SO double-mode Cepheids}%7.4

Periodic amplitude and phase modulation in FO/SO double-mode
Cepheids is by no mean a rare phenomenon. According to our
analysis, its incidence rate is at least $18.7\pm 3.8$\%. The true
rate can be perhaps even as high as 35\%, if FO/SO-PC variables
are confirmed to undergo Blazhko modulation with very long
periods, as we suspect.

In Fig.\thinspace 17 we present the Petersen diagram for our
sample of FO/SO double-mode Cepheids of the LMC. Blazhko FO/SO
pulsators and FO/SO-PC pulsators are marked with filled and open
circles, respectively. Blazhko FO/SO variables are not detected at
all for $P_1 < 0.6$\thinspace day, but for longer periods they
intermingle with "normal" double-mode Cepheids. The same can be
said about FO/SO-PC variables, except that the short period cutoff
is at $P_1 = 0.5$\thinspace day. Apart from amplitude and phase
variability, Blazhko pulsators and PC pulsators do not differ from
the rest of FO/SO double-mode Cepheids.

Fig.\thinspace 17 shows, however, that the incidence rate of
Blazhko effect might depend on pulsation period. To address this
point we display in Fig.\thinspace 18 the histogram of first
overtone periods for our sample of FO/SO Cepheids. The incidence
rate of Blazhko effect peaks at 0.8\thinspace day~$< P_1 <
1.0$\thinspace day, where it reaches 48\%. It sharply declines
below 15\% for both longer and shorted periods. While the decline
of the incidence rate towards shorted periods ({\it
i.e.}\thinspace lower luminosities) can be attributed to the
selection effects, the decline towards longer periods must be
real.

\subsection{What Causes the Modulation ?}%7.5

Any model of Blazhko FO/SO double-mode Cepheids has to explain two
most basic properties of these stars:

\begin{itemize}
\item both radial modes are modulated with the same period.
\item the amplitude variations of the two modes are anticorrelated: maximum
      of one amplitude coincides with minimum of the other.
\end{itemize}

\noindent Many different ideas have been put forward to explain
Blazhko modulation in monoperiodic RR~Lyrae stars (see Stothers 2006
for most recent summary). Two primary contenders, which are most
popular nowadays, are the oblique magnetic pulsator model
(Shibahashi 1995, 2000) and the 1:1 resonance model (Nowakowski \&
Dziembowski 2001). Another, very different scenario has been
proposed recently by Stothers (2006). We will argue, that all these
models fail to account for modulation observed in the FO/SO
double-mode Cepheids.

Shibahashi's oblique magnetic pulsator model assumes presence of a
dipole magnetic field inclined to the rotation axis of the star.
The field introduces quadru\-pole distortion to the radial mode.
As the star rotates, the distortion is viewed from different
aspect angles, which leads to variation of an apparent pulsation
amplitude. Somewhat similar idea has been considered by Kov\'acs
(1995; see also Bal\'azs 1959). In his scenario the distortion is
caused by an $\ell=1$ nonradial mode, which is aligned with a
magnetic axis of the star. The nonradial mode is excited through a
1:1 resonance with the radial pulsation. Consequently, frequencies
of both modes are exactly synchronized. Both in Shibahashi's and
in Kov\'acs's models physical amplitudes of pulsation are
constant. Modulation of the {\it observed} amplitudes is caused by
geometrical effect of rotation. The models naturally explain why
both radial modes in FO/SO Cepheid are modulated with the same
period. However, in the above scenario both modes should reach
maximum amplitudes simultaneously, because they are both distorted
in the same way. This is not what we observe. Thus, the Blazhko
effect in FO/SO double-mode Cepheids cannot be reproduced by the
oblique magnetic pulsator.

The model proposed by Nowakowski \& Dziembowski (2001, see also
Kov\'acs 1995) assumes a 1:1 resonant coupling of the radial mode
to a {\it pair of nonradial modes} of $\ell=1, m=\pm 1$. Such a
mechanism generates a triplet of equally spaced frequencies. The
physical amplitudes of the modes are constant, but beating of
their equidistant frequencies leads to a slow periodic modulation
of apparent amplitude and phase of pulsation. The 1:1 resonance
model naturally explains triplets observed in frequency spectra of
Blazhko FO/SO Cepheids. Quintuplets can also be easily understood,
by generalizing the model to coupling with modes of $\ell=2$.
However, in the scenario of Nowakowski \& Dziembowski (2001)
modulation of each radial mode is an independent process.
Therefore, there is no reason why both radial modes should be
modulated with {\it exactly} the same period (although modulation
periods shouldn't be very different). There is also no reason why
amplitude variations of the two modes should be in any particular
phase relation to each other, as is observed. Finally, the 1:1
resonance model does not explain why in vast majority of cases
modulation is observed either for both radial modes or for none of
them.

The new model (or rather an idea) of Stothers (2006) fails to
explain the observations of Blazhko FO/SO Cepheids, as well. In
his scenario it is assumed that turbulent convection in the
stellar envelope is cyclically weakened and strengthened by a
transient magnetic field generated by dynamo mechanism. As
convection has a strong effect on the amplitudes of pulsation
({\it e.g.}\thinspace Feuchtinger 1999), the latter become
modulated, too. In this picture we would expect both modes in a
FO/SO Cepheid to reach maximum amplitudes at the same time,
because both amplitudes vary in response to a common factor -- a
changing strength of convection. This prediction is in obvious
conflict with observations.

At this stage, the mechanism causing amplitude and phase
modulation of FO/SO double-mode Cepheids remains unknown.
Nevertheless, the available observations already provide some
important clues. The fact that the two modes are always modulated
with the same period proves that their behaviour is not
independent. Both modes must be part of the same dynamical system.
The fact that high amplitude of one mode always coincides with low
amplitude of the other mode strongly suggests that energy transfer
between the two modes is involved. Thus, available evidences point
towards some form of mode interaction, which causes a periodic
amplitude and phase modulation of the modes. Such a modulation
(limit cycle) can be induced by a resonant coupling of a radial
mode with another radial or nonradial mode (Moskalik 1986;
Van~Hoolst 1995). Because of the nonlinear frequency
synchronization, the resonant mode {\it will not appear} as an
independent peak in the power spectrum. It is enough that only one
of the observed radial modes is modulated by the resonant
coupling. This modulation will be carried over to the other radial
mode through the so-called cross-saturation effect. Speaking in
physical term, the two radial modes compete for the same driving
($\kappa$ mechanism in the He$^{+}$ ionization zone). When the
amplitude of one mode is suppressed by whatever mechanism ({\it
e.g} by a resonance), it allows the other mode to grow. This
explains in a very simple and natural way why the two radial modes
are modulated with the common period and why their amplitudes are
always anticorrelated.

\section{Conclusions}%8

Taking advantage of a large and homogenous photometric dataset
collected during OGLE-II project, we have performed a systematic
frequency analysis of classical Cepheids identified in the Large
Magellanic Cloud (Udalski {\it et~al.}\thinspace 1999b;
Soszy\'nski {\it et~al.}\thinspace 2000). This study has been
aimed at finding multiperiodic and nonstationary variables among
Pop.\thinspace I Cepheids and at establishing reliable statistics
of different forms of their pulsational behaviour. The main body
of our survey has been conducted with OGLE-II photometry, but in
cases when frequency resolution turned out to be insufficient, we
have resorted to photometry collected by the MACHO experiment.

We have discovered two {\it triple-mode} Cepheids, which pulsate
with three lowest radial overtones simultaneously excited.
Triple-mode radial pulsators have been known before, but only
among High Amplitude $\delta$~Scuti stars (Kov\'acs \& Buchler
1994; Antipin 1997b; Handler, Pikall \& Diethelm 1998; Wils {\it
et~al.}\thinspace 2008). This is the first detection of such
pulsators among Cepheids. Three additional triple-mode Cepheids
have been identified in the LMC by Soszy\'nski\thinspace {\it
et~al.} (2008a). These rare variables are a very valuable trophy,
since their seismological analysis can strongly constrain the
stellar evolution theory (Moskalik \& Dziembowski 2005).

In 19\% of FO/SO double-mode Cepheids we have detected {\it
periodic variability of amplitudes and phases} of the two radial
modes. Another 16\% of FO/SO pulsators is suspected of undergoing
the same type of modulation, but on timescales longer than our
data. This phenomenon is analogous to the Blazhko modulation in
the RR~Lyr stars, which was first discovered a century ago
(Blazhko 1907). In Blazhko FO/SO Cepheids both modes are modulated
with a common period, always in excess of 700\thinspace days. The
amplitudes of the modes vary in opposite phases, that is a maximum
amplitude of one mode coincides with minimum amplitude of the
other. In frequency domain, the modulation manifests itself as
splitting of each radial mode into a triplet or sometimes a
quintuplet of equidistant peaks, spaced by frequency of the
modulation.

The discovery of modulated FO/SO Cepheids shows that Blazhko
effect and double-mode pulsations are not mutually exclusive.
These stars will play a very important role in selecting a proper
model to explain the Blazhko phenomenon. Observations of {\it two}
modulated modes in one star will put any proposed scenario to a
very stringent test. Two most popular models, the oblique magnetic
pulsator model (Shibahashi 2000) and the 1:1 resonance model
(Nowakowski \& Dziembowski 2001) have already failed this test and
should be ruled out, at least in the case of FO/SO Cepheids.

Perhaps the most exciting result of our survey is detection of
{\it nonradial modes} in classical Cepheids. Such modes have been
found in 42 overtone pulsators (which constitutes 9\% of the
sample) and in three FU/FO double-mode pulsators. They have {\it
not been found} in Cepheids pulsating in the fundamental mode.

We find two different types of nonradial modes in Cepheids. In
most cases these modes are detected very close to the first radial
overtone, but in several stars they appear at considerably higher
frequencies. In the later case, the measured period ratio falls in
a narrow range of $0.60-0.64$, which places the nonradial mode in
the close vicinity of the (unobserved) fourth radial overtone. The
observed frequency pattern cannot be explained by any form of
amplitude or phase modulation, it is also incompatible with
excitation of radial modes. This makes interpretation in terms of
nonradial modes the only possibility.

Detection of nonradial modes in classical Cepheids has been
claimed before by Kovtyukh {\it et~al.}\thinspace (2003), based on
observations of bumps in spectral line profiles of four galactic
Cepheids. Such structures can be indicative of nonradial
oscillations ({\it e.g.}\thinspace Telting 2003). However, the
authors have not shown that the observed line profiles vary in a
periodic way, or that they vary with their own period different
from that of the radial mode. This is an important test, because
bumps in the spectral lines can be caused by mechanisms other than
nonradial pulsations ({\it e.g.}\thinspace by shocks). Therefore,
observations presented by Kovtyukh {\it et~al.}\thinspace (2003)
are in our opinion inconclusive. The results of frequency analysis
discussed in the current paper provide the first solid evidence of
nonradial modes excitation in classical Cepheids.

Discovery of nonradial modes in Cepheids poses a major challenge
to the stellar pulsation theory. In order to be photometrically
detectable, nonradial modes have to be of low spherical degree,
$\ell$ (Dziembowski 1977). However, linear pulsation calculations
predict, that in classical Cepheids all modes of $\ell < 4$ should
be heavily damped (Osaki 1977). This conclusion has recently been
confirmed by new calculations of Mulet-Marquis {\it
et~al.}\thinspace (2007), who have found that all nonradial modes
of $\ell < 5$ should be damped. We note, that this is a very
different situation from that encountered in the RR~Lyr models,
where many modes of $\ell=1,2$ are linearly excited (Dziembowski
\& Cassisi 1999).

When writing of this manuscript was almost completed, we learnt
about a new preprint on LMC Cepheids by Soszy\'nski {\it
et~al.}\thinspace (2008b). The authors have analyzed data
collected during the third phase of the OGLE project (OGLE-III).
Among many interesting results, they have also detected secondary
periodicities in all subclasses of Pop.\thinspace I Cepheids. This
provides an important confirmation of our results with independent
data. The incidence rates given by Soszy\'nski {\it
et~al.}\thinspace are very similar to ours in case of the
double-mode pulsators, but significantly higher than ours in case
of the FO Cepheids. This is most likely due to much longer time
span of their photometry and higher number of measurements.
Interestingly, Soszy\'nski {\it et~al.}\thinspace has detected
secondary periodicities also in the FU Cepheids, which we have
not. A detailed comparison of incidence rates derived by us and by
Soszy\'nski {\it et~al.}\thinspace is beyond the scope of the
present paper. We only note, that such comparison requires some
caution, because the latter authors do not discriminate between
se\-condary periodicities resolved from the primary frequency and
those, which are not resolved.

\Acknow{This work was supported in part by MNiSW Grant No. 1~P03D~011~30.}

\newpage

\begin{figure}%1
\includegraphics[width=18.8cm]{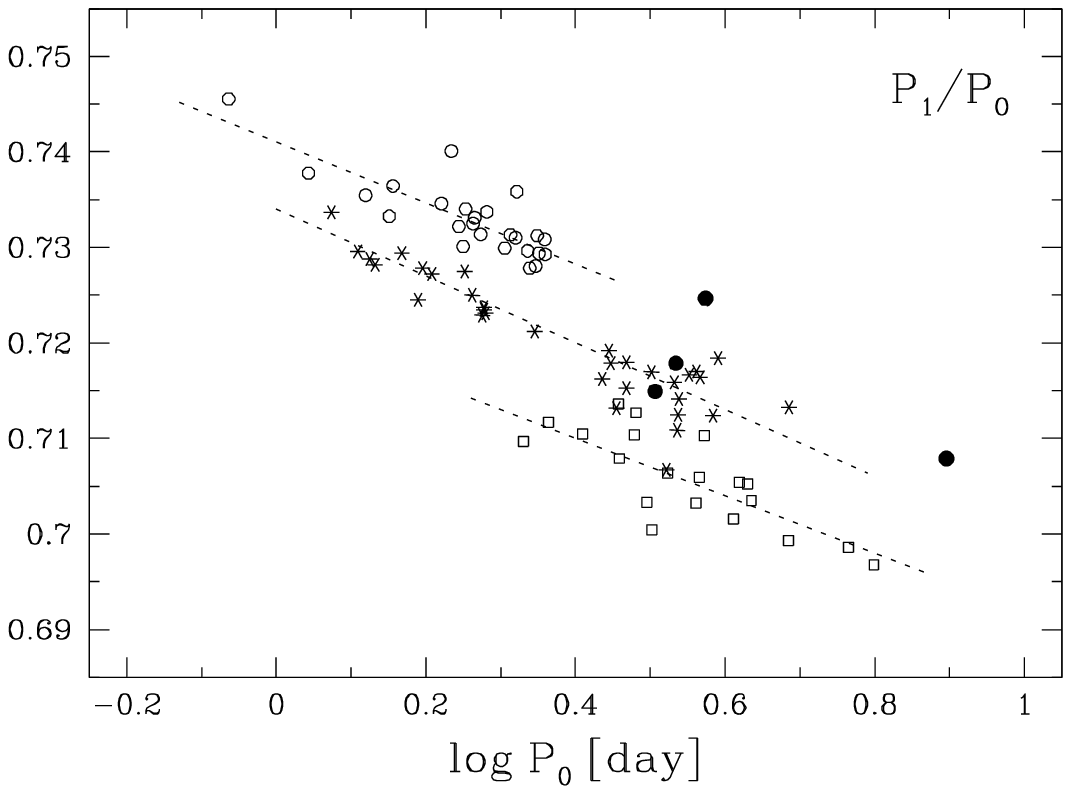}
\vskip -309pt
\FigCap{Petersen diagram for FU/FO double-mode Cepheids in the Galaxy (open
        squares), LMC (asterisks) and SMC (open circles). New LMC double-mode
        Cepheids are displayed with filled circles. Dotted lines represent the
        best linear fits to the period ratios observed in each population.}
\end{figure}

\vskip 50pt

\begin{figure}%2
\includegraphics[width=18.8cm]{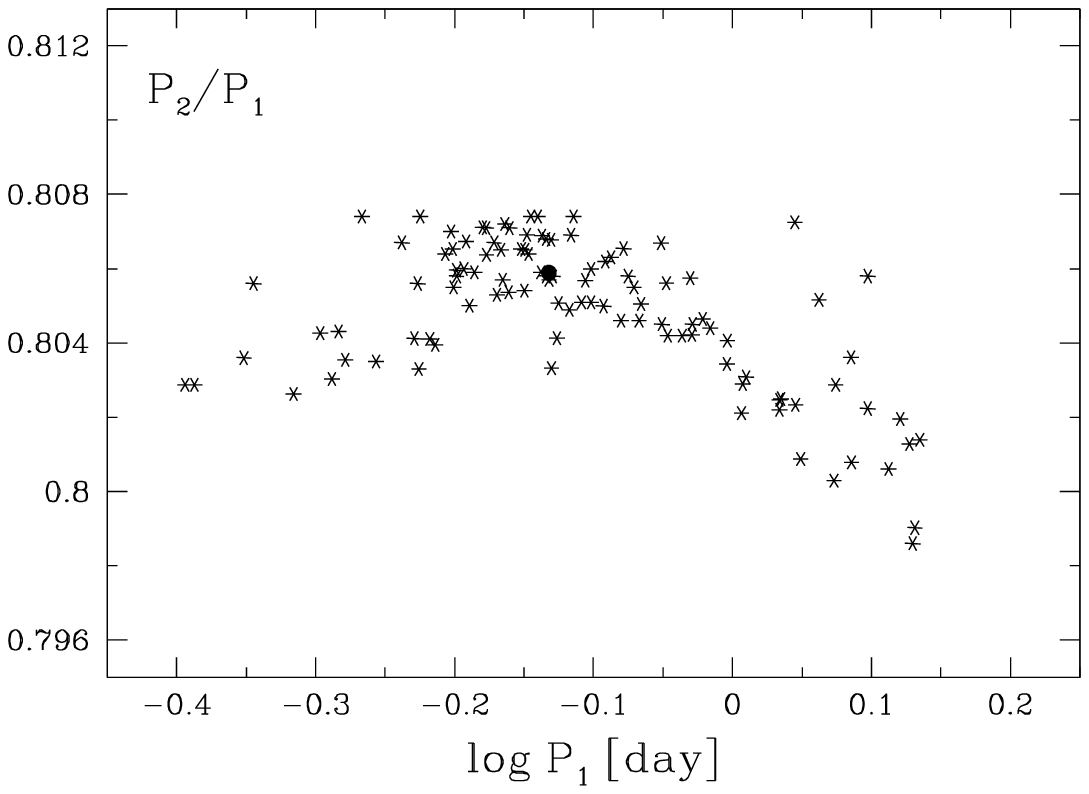}
\vskip -309pt
\FigCap{Petersen diagram for FO/SO double-mode Cepheids in the LMC. New
        double-mode Cepheid SC20--112788 displayed with a filled circle.}
\end{figure}

\begin{figure}%3
\hskip -52pt
\includegraphics[width=17.9cm]{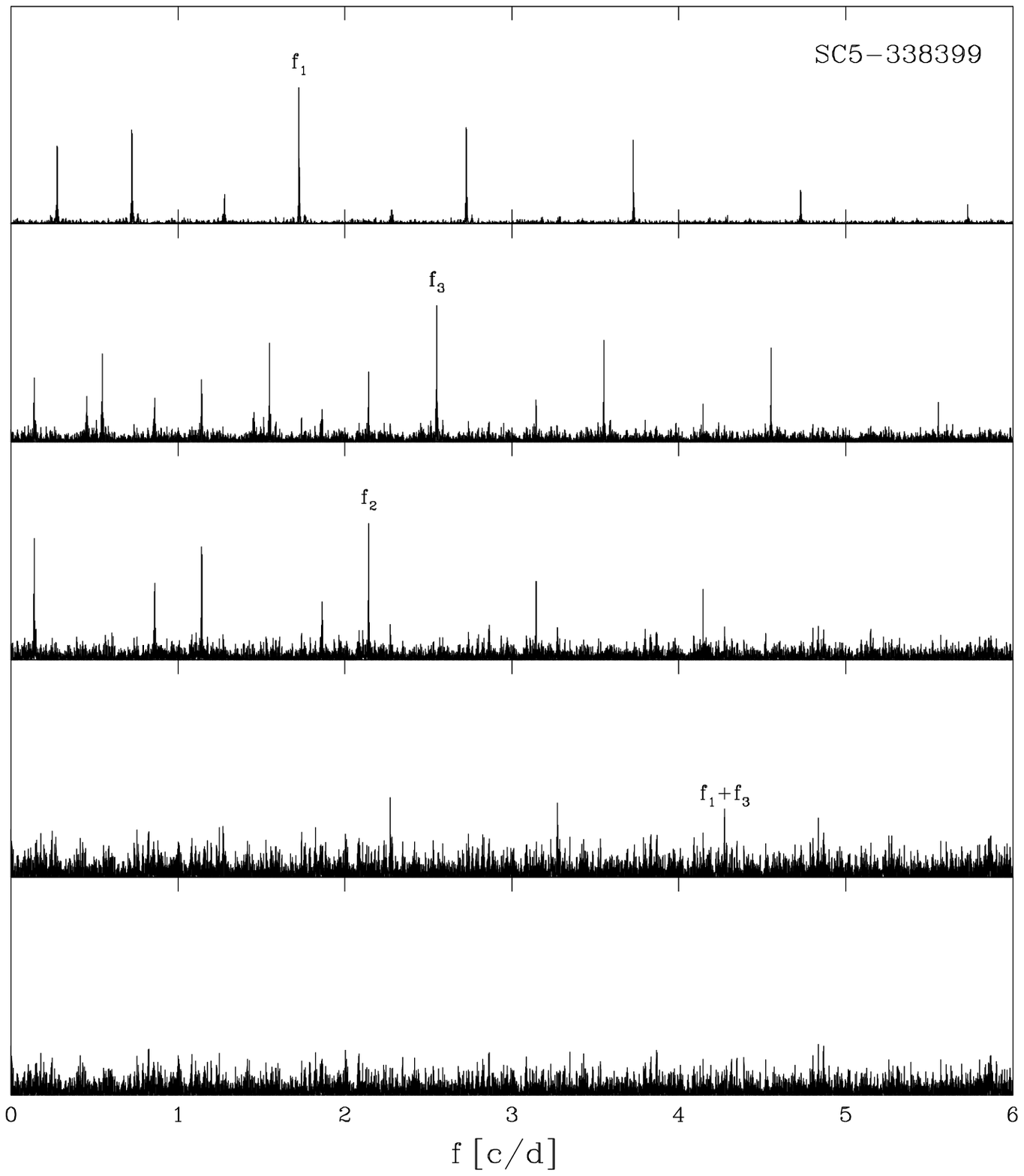}
\vskip -75pt
\FigCap{Prewhitening sequence for triple-mode Cepheid SC5--338399. Upper panel
        displays power spectrum of original data. Lower panels show power
        spectra after removing consecutive frequencies.}
\end{figure}

\begin{figure}%4
\includegraphics[width=18.8cm]{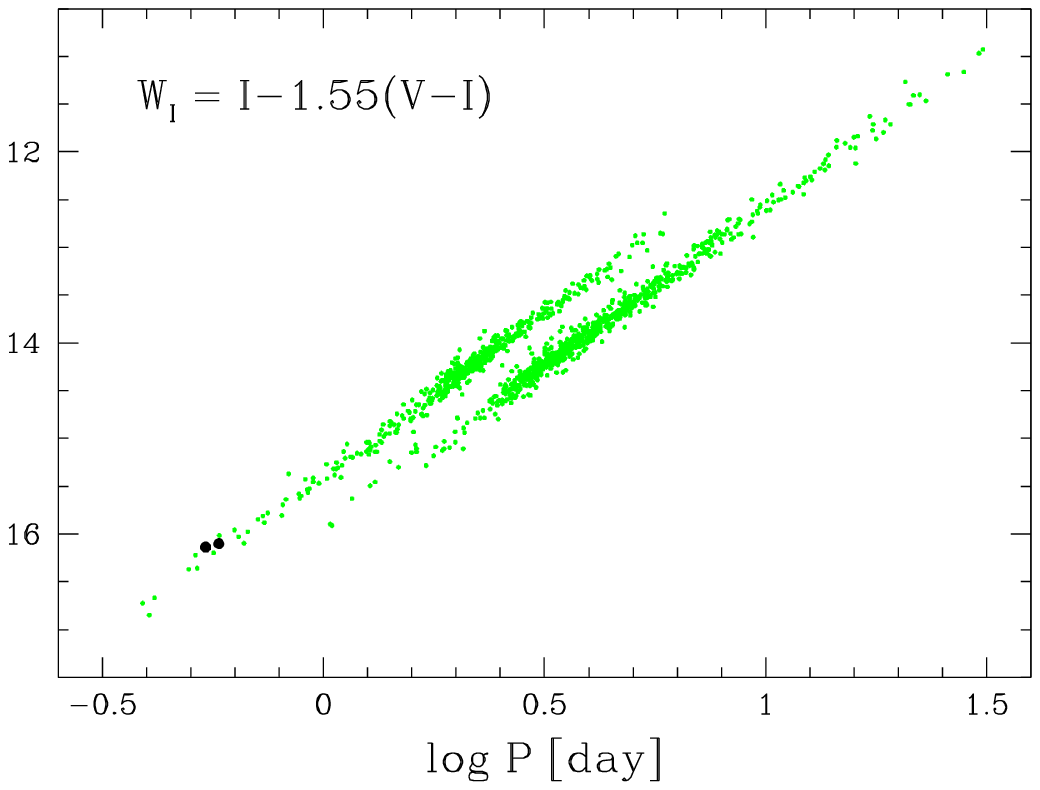}
\vskip -309pt
\FigCap{$W_I$ index $P-L$ relations for LMC Cepheids. Upper and lower
        sequences correspond to first overtone and fundamental mode Cepheids,
        respectively. First overtone periods of triple-mode Cepheids
        SC3--360128 and SC5--338399 are displayed with black symbols.}
\end{figure}

\begin{figure}%5
\hskip -52pt
\includegraphics[width=17.9cm]{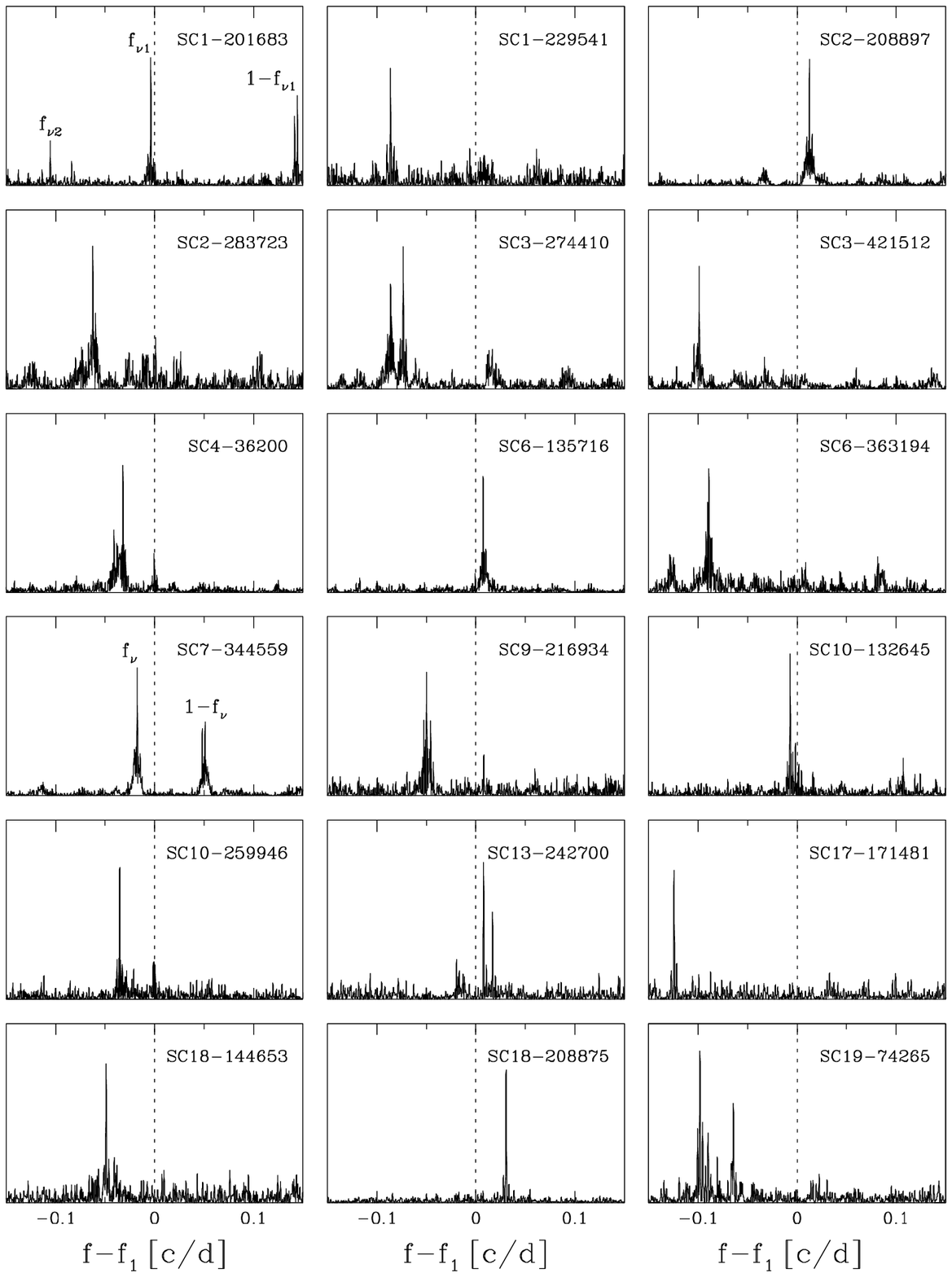}
\vskip -2pt
\FigCap{Prewhitened power spectra of sample of FO-$\nu$ Cepheids with small
        $|\Delta f|$. Frequencies of removed radial modes indicated by dashed
        lines.}
\end{figure}

\begin{figure}%6
\hskip -51pt
\includegraphics[width=17.7cm]{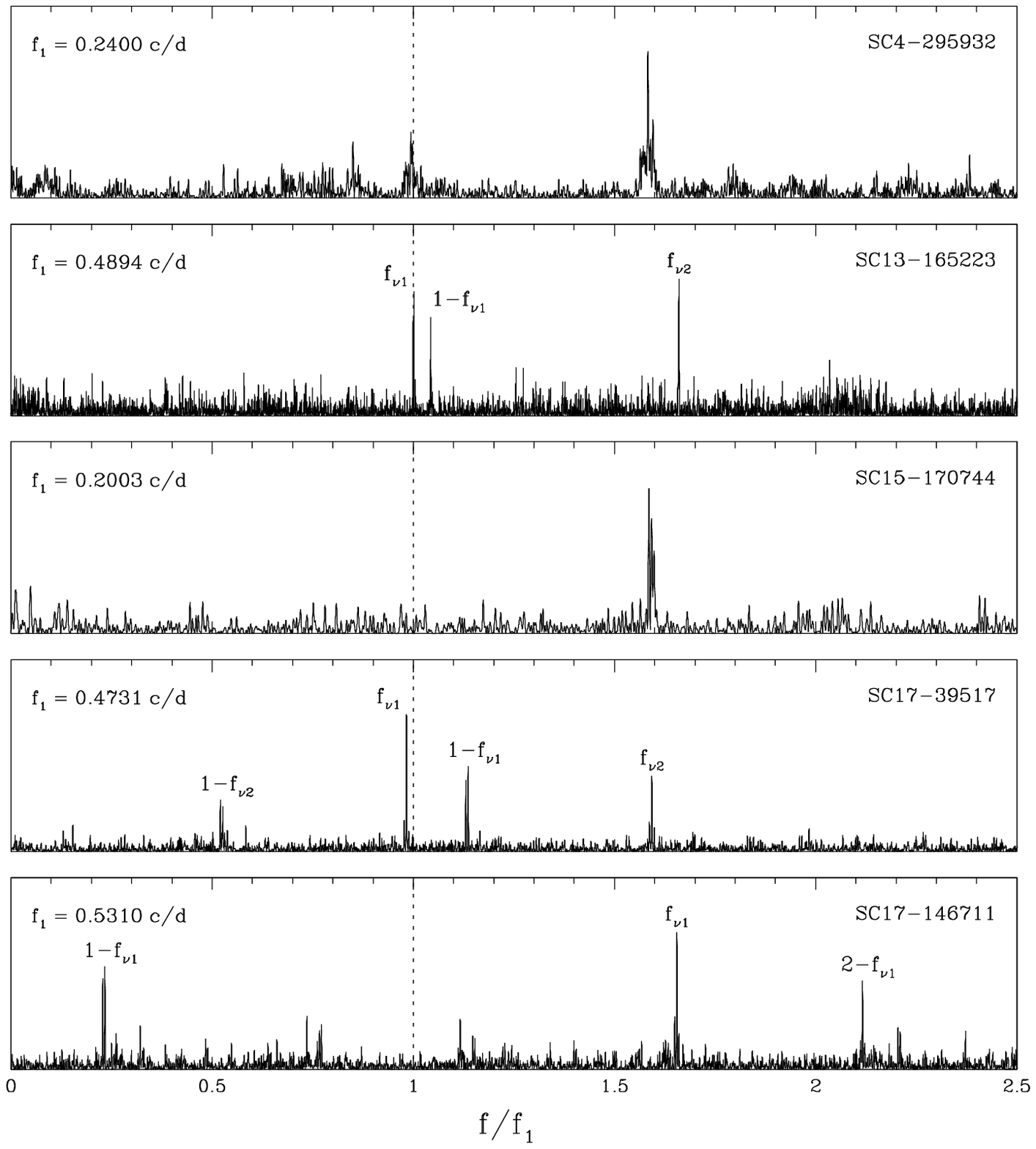}
\vskip -2pt
\FigCap{Prewhitened power spectra of sample of FO-$\nu$ Cepheids with
        $P_{\nu}/P_1 = 0.60 - 0.64$. Frequencies of removed radial modes
        indicated by dashed lines.}
\end{figure}

\begin{figure}%7
\includegraphics[width=18.8cm]{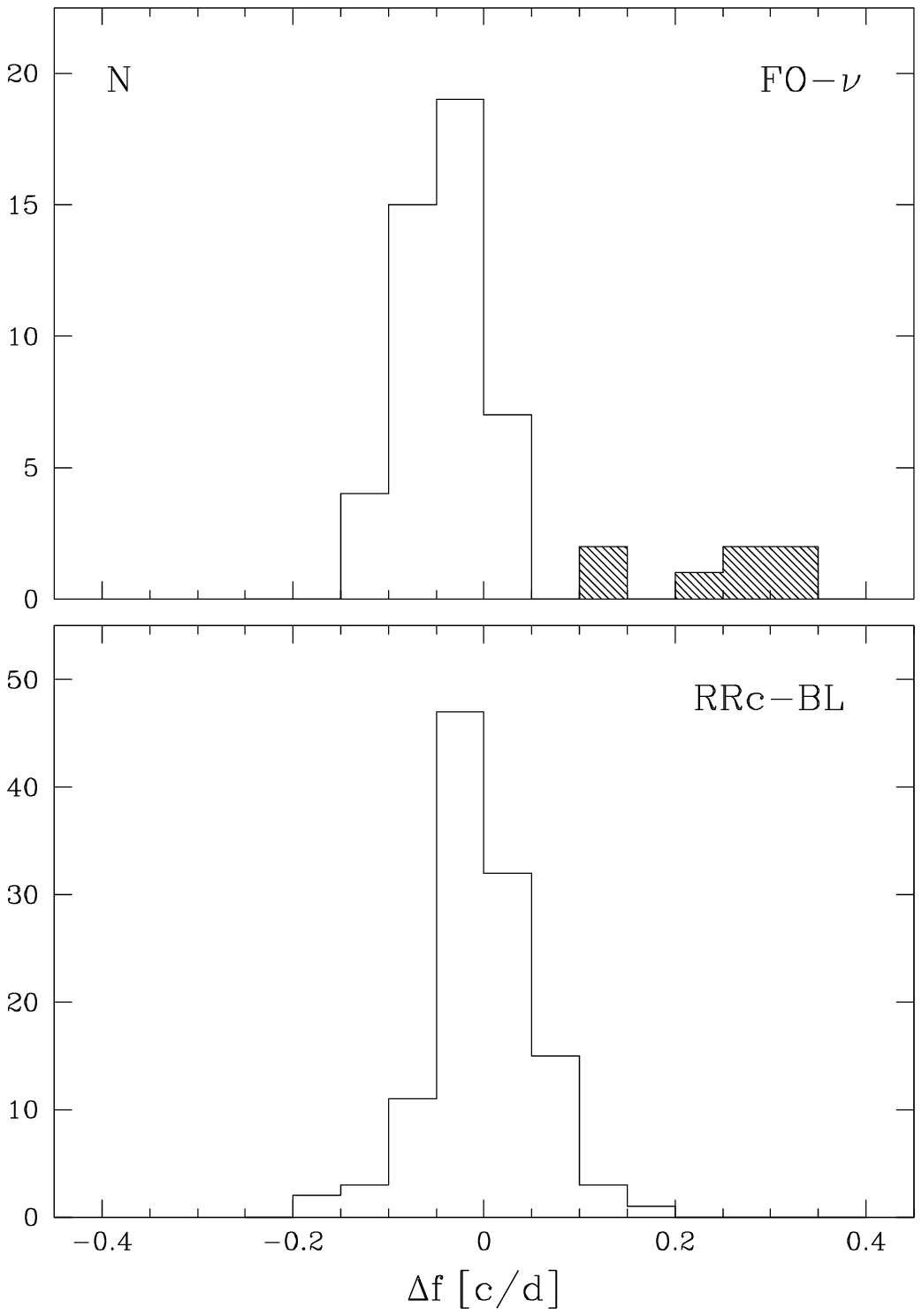}
\vskip -121pt
\FigCap{Distribution of frequency differences $\Delta f = f_{\nu} - f_1$ for
        LMC FO-$\nu$ Cepheids and for LMC RRc-BL stars (from Nagy \& Kov\'acs
        2006). For RRc-BL variables with frequency triplets, only higher
        side peaks are taken into account. Shaded area marks variables with
        $P_{\nu}/P_1 = 0.60 - 0.64$.}
\end{figure}

\begin{figure}%8
\hskip -35pt
\includegraphics[width=17.2cm]{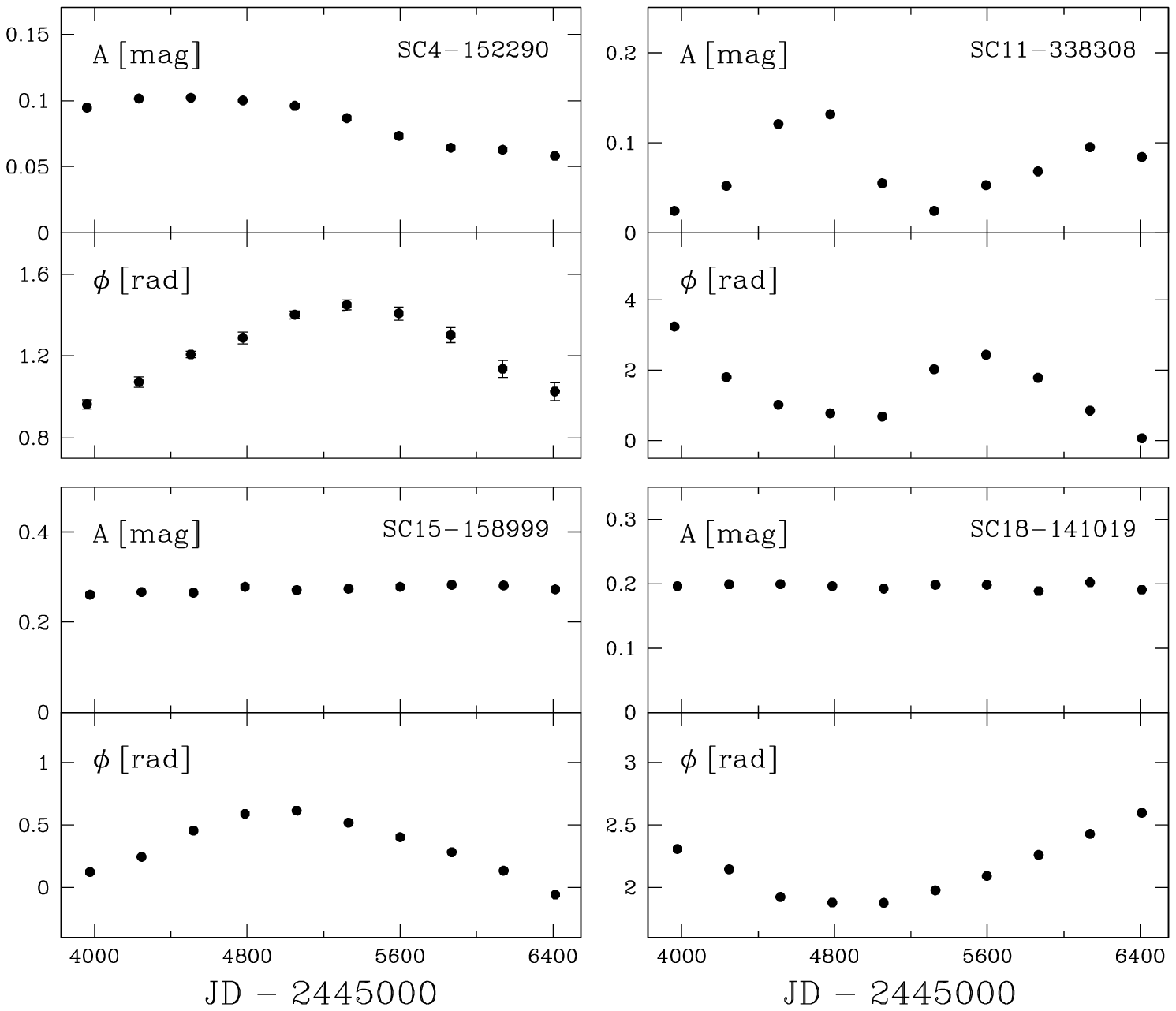}
\vskip -156pt
\FigCap{Amplitude and phase variations of sample of FO-PC Cepheids.
        $\pm 1\sigma$ error bars shown when larger then the symbols.
        In SC4--152290 and SC11--338308 both amplitudes and phases change in
        time. In SC15--158999 and SC18--141019 amplitudes are constant and only
        pulsation phases vary.}
\end{figure}

\begin{figure}%9
\includegraphics[width=18.8cm]{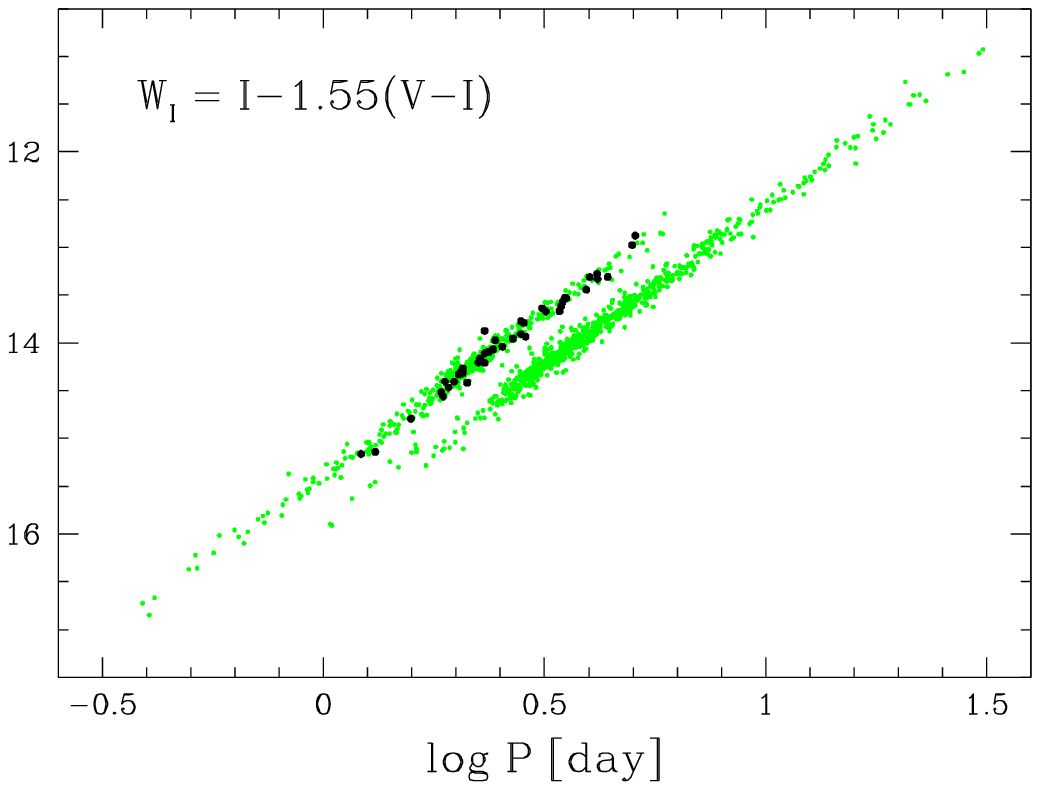}
\vskip -309pt
\FigCap{$W_I$ index $P-L$ diagram for LMC Cepheids. Dominant (radial) periods
        of FO-$\nu$ Cepheids are displayed with black symbols.}
\end{figure}

\begin{figure}%10
\includegraphics[width=18.8cm]{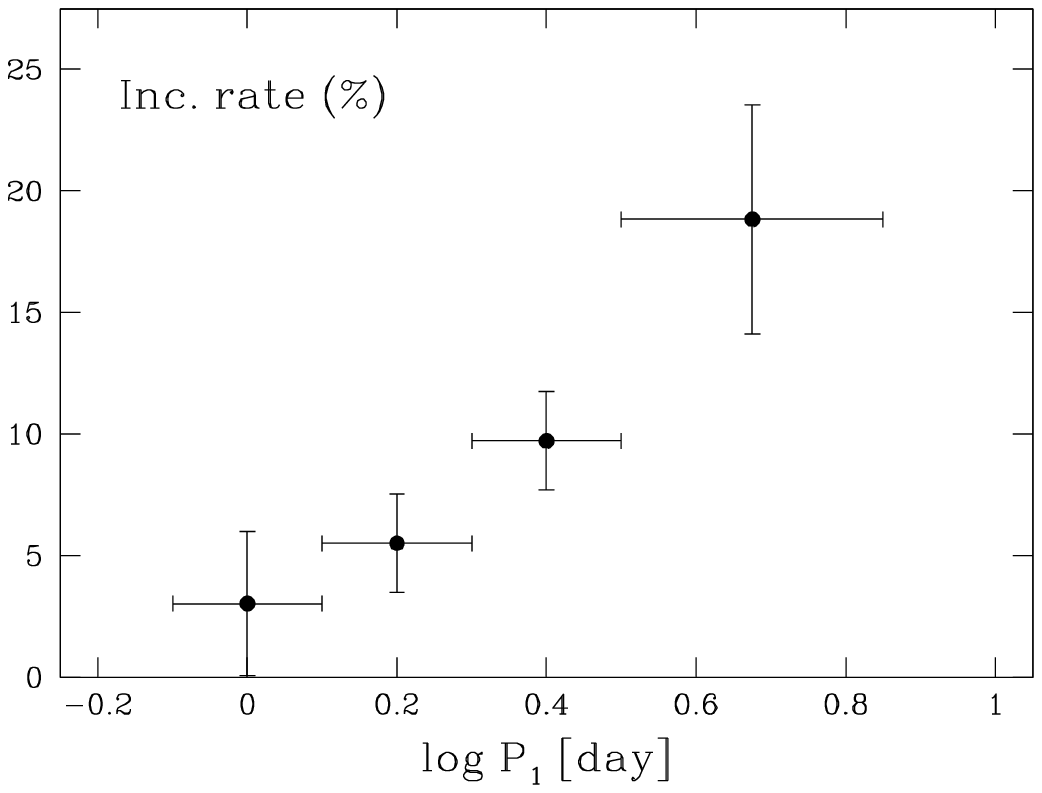}
\vskip -309pt
\FigCap{Incidence rate of LMC FO-$\nu$ Cepheids {\it vs.} first overtone
        period.}
\end{figure}

\begin{figure}%11
\hskip -52pt
\includegraphics[width=17.9cm]{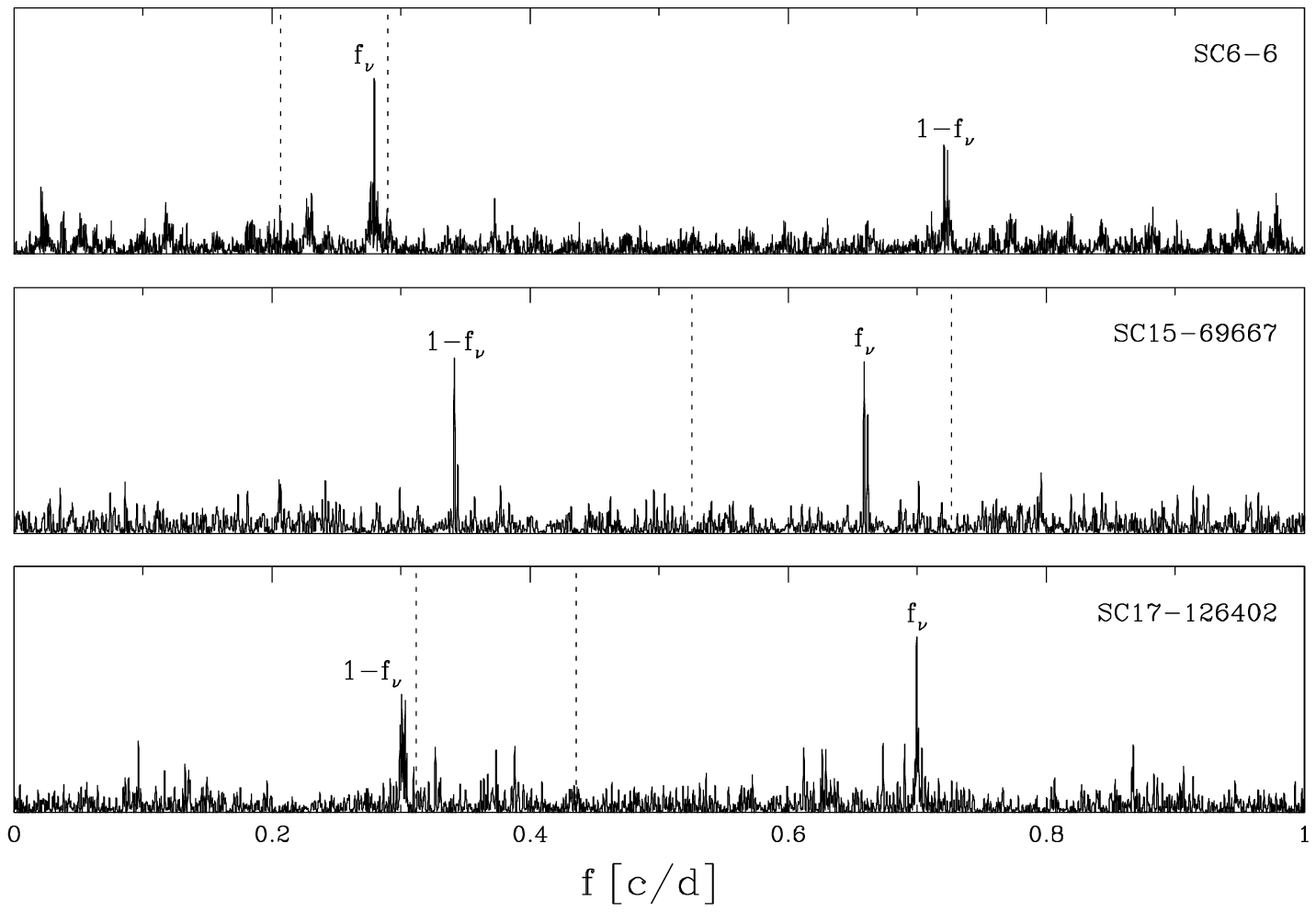}
\vskip -159pt
\FigCap{Power spectra of FU/FO-$\nu$ double-mode Cepheids after prewhitening
        with frequencies of the two radial modes and their linear combinations.
        Removed radial frequencies indicated by dashed lines.}
\end{figure}

\begin{figure}%12
\hskip -72pt
\includegraphics[width=16.6cm]{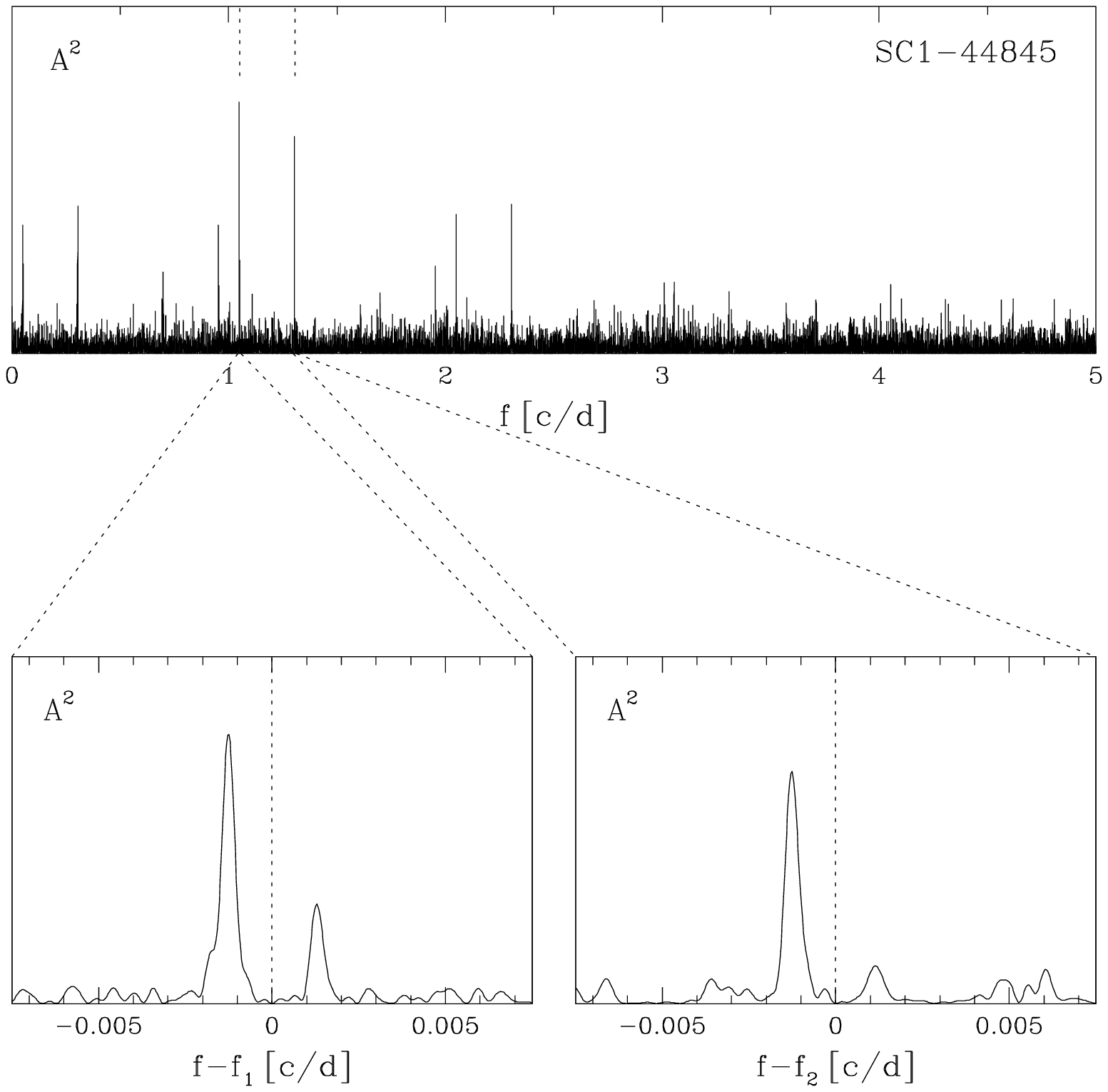}
\vskip -87pt
\FigCap{Prewhitened power spectrum of Blazhko FO/SO double-mode Cepheid
        SC1--44845. Frequencies of removed radial modes indicated by dashed
        lines. Lower panels display the fine structure around the radial
        modes.}
\end{figure}

\begin{figure}%13
\hskip -54pt
\includegraphics[width=18cm]{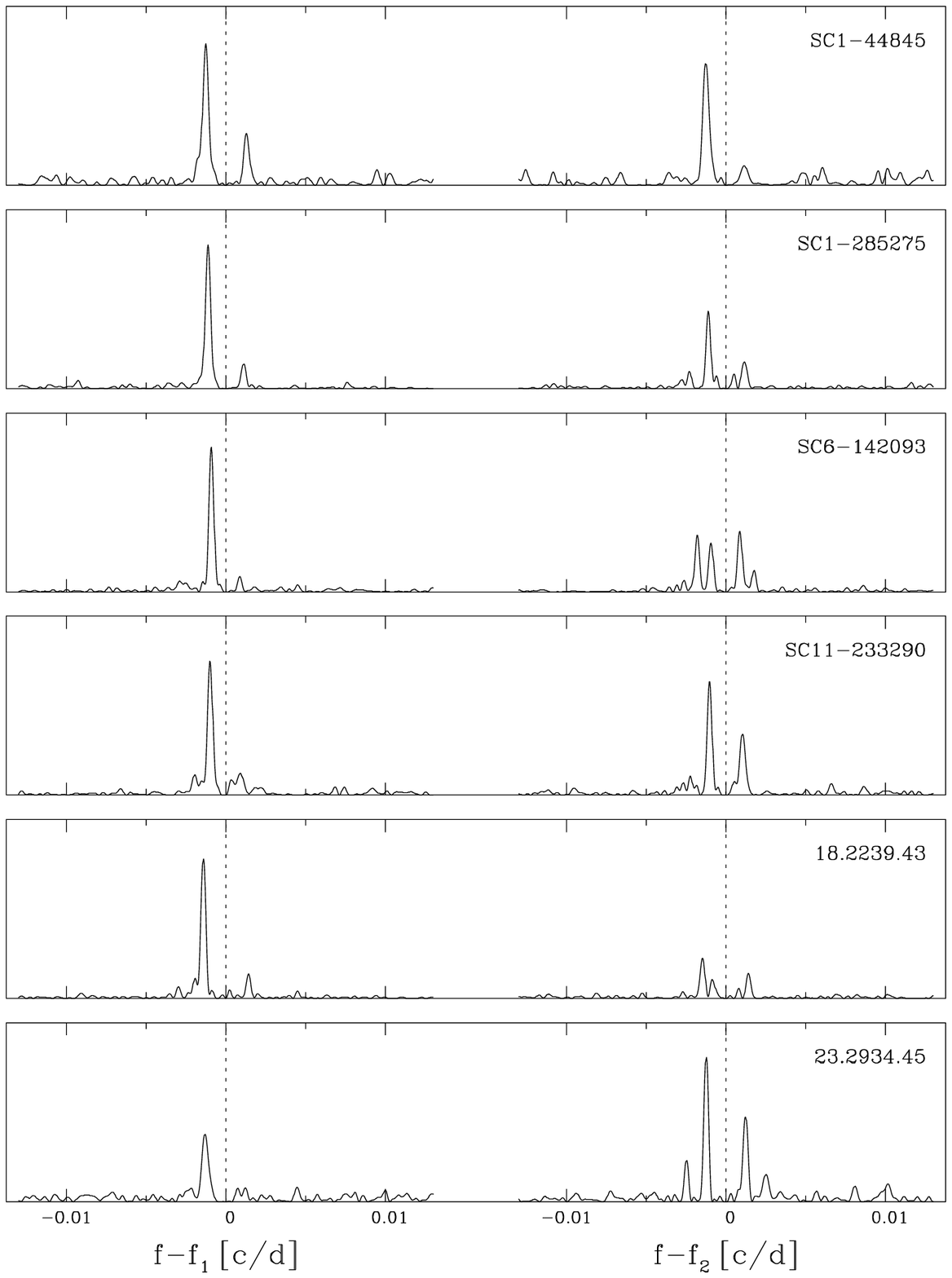}
\vskip -2pt
\FigCap{Prewhitened power spectra of sample of Blazhko FO/SO double-mode
        Cepheids. Frequencies of removed radial modes indicated by dashed
        lines.}
\end{figure}

\begin{figure}%14
\includegraphics[width=18.8cm]{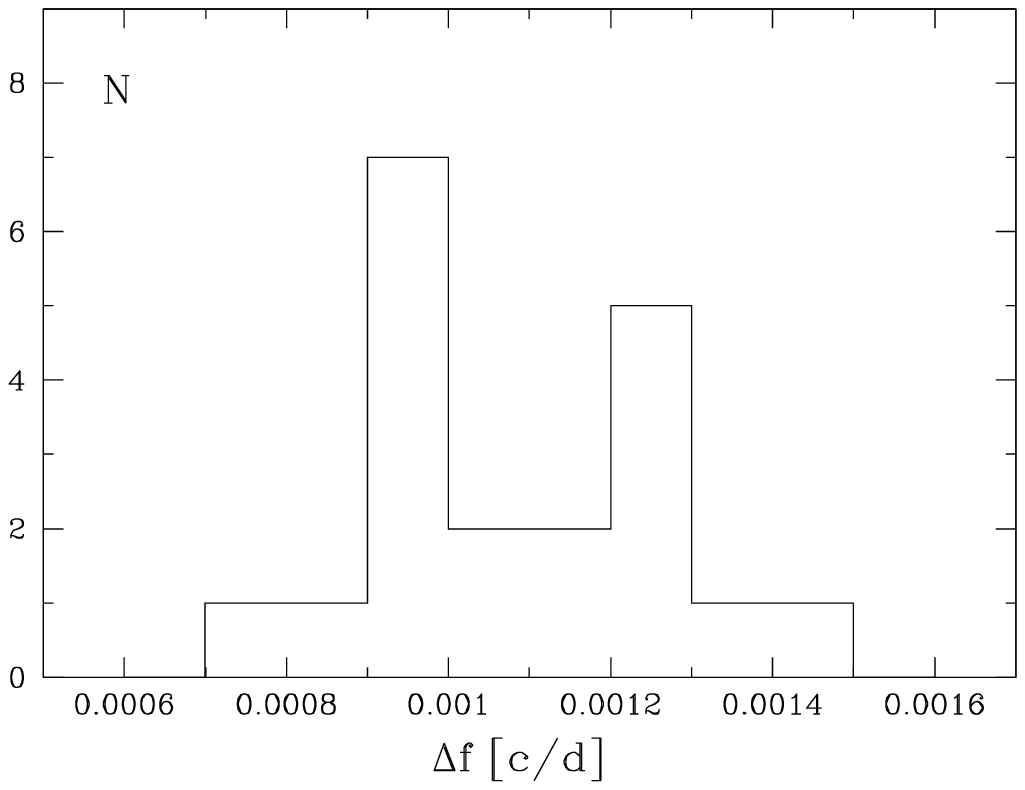}
\vskip -309pt
\FigCap{Distribution of modulation frequencies, $\Delta f$, for Blazhko FO/SO
        double-mode Cepheids.}
\end{figure}

\begin{figure}%15
\hskip -31pt
\includegraphics[width=17.1cm]{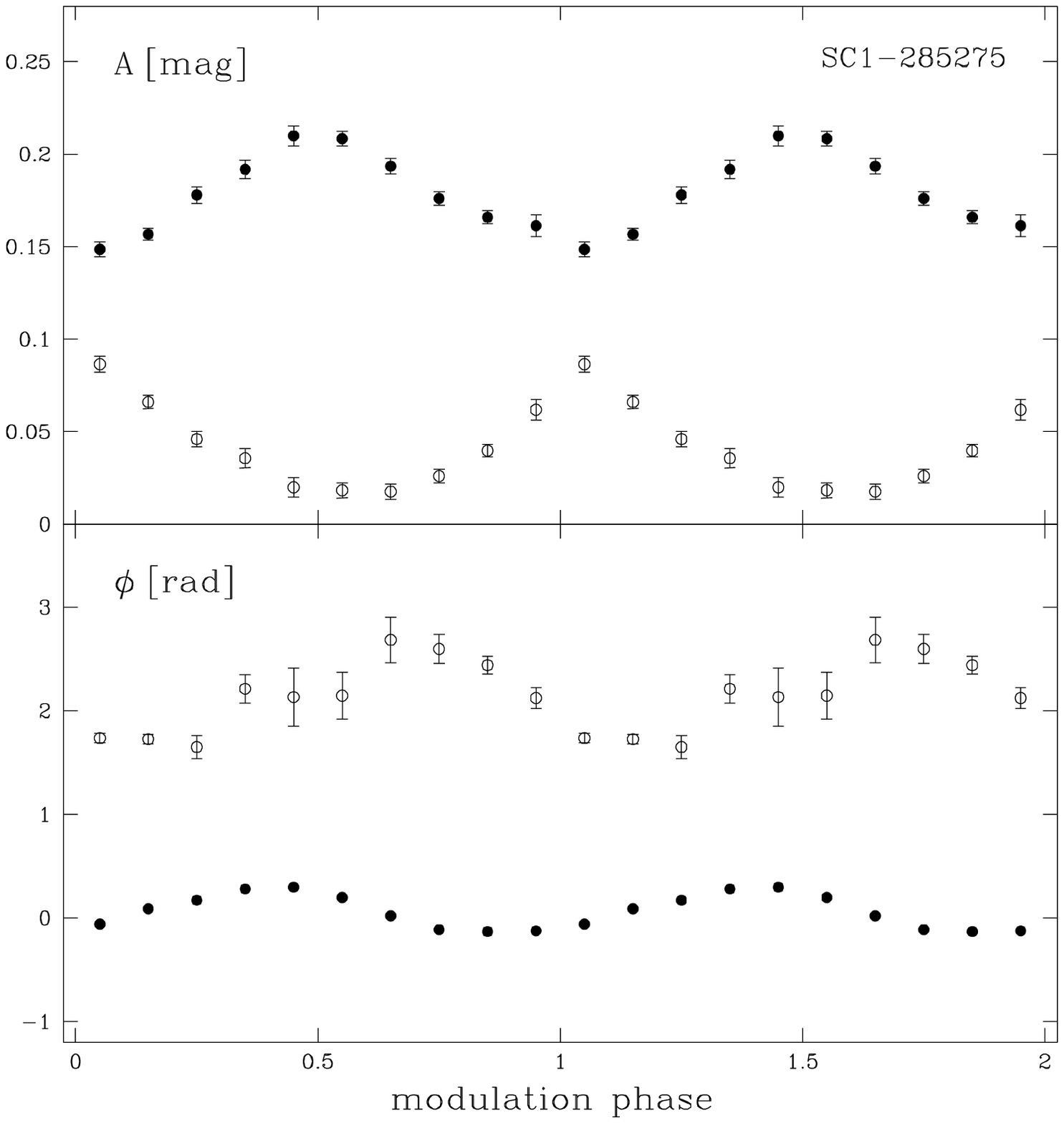}
\vskip -86pt
\FigCap{Periodic modulation of amplitudes and phases of radial modes in FO/SO
        double-mode Cepheid SC1--285275. First and second overtones are
        displayed with filled and open circles, respectively. The error bars
        are $\pm 1\sigma$. Modulation period is $P_B = 891.6$\thinspace day.}
\end{figure}

\begin{figure}%16
\hskip -32pt
\includegraphics[width=17.1cm]{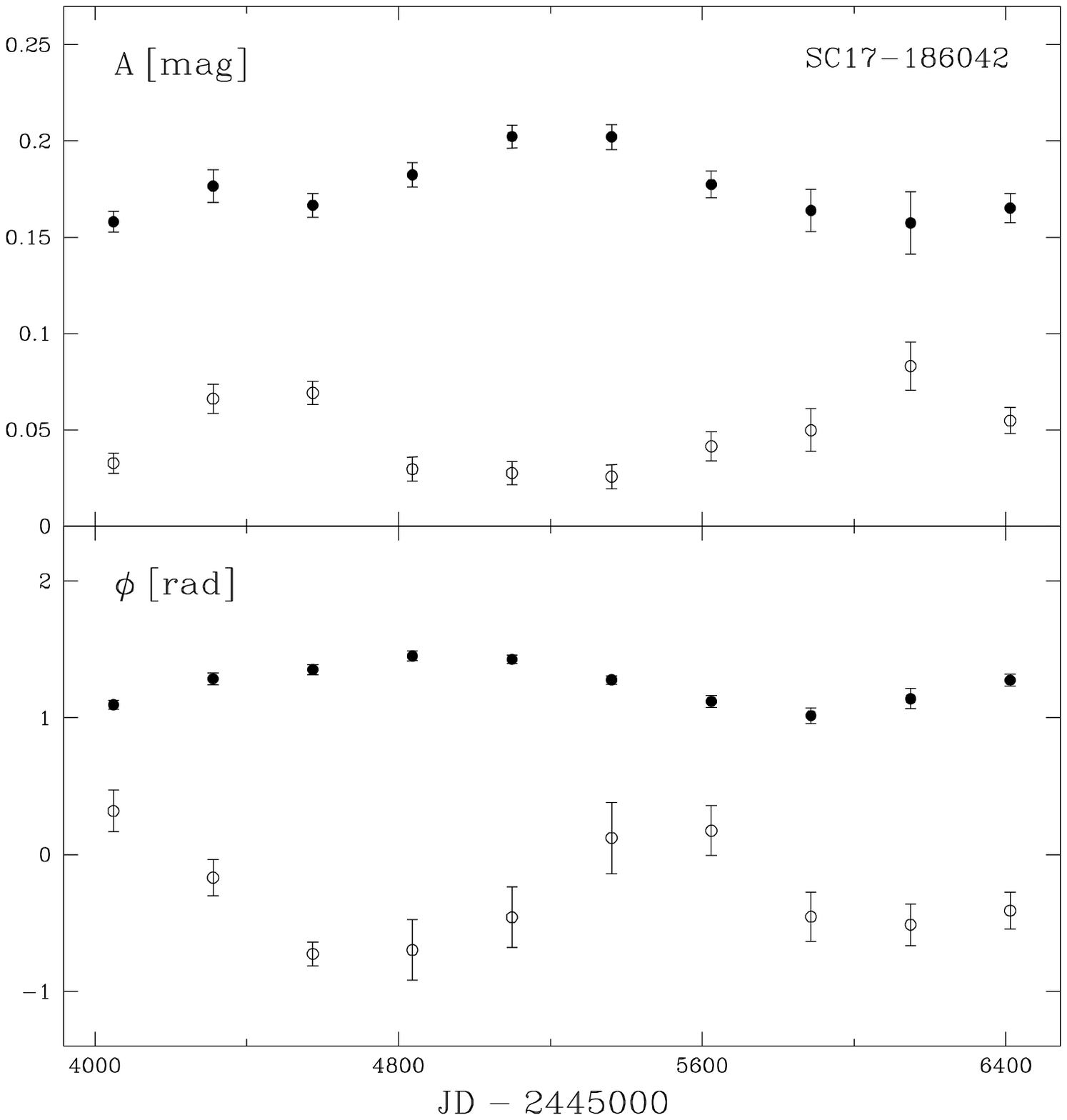}
\vskip -88pt
\FigCap{Amplitude and phase variations in FO/SO-PC double-mode Cepheid
        SC17--186042. Symbols are the same as in Fig.\thinspace 15.}
\end{figure}

\begin{figure}%17
\includegraphics[width=18.8cm]{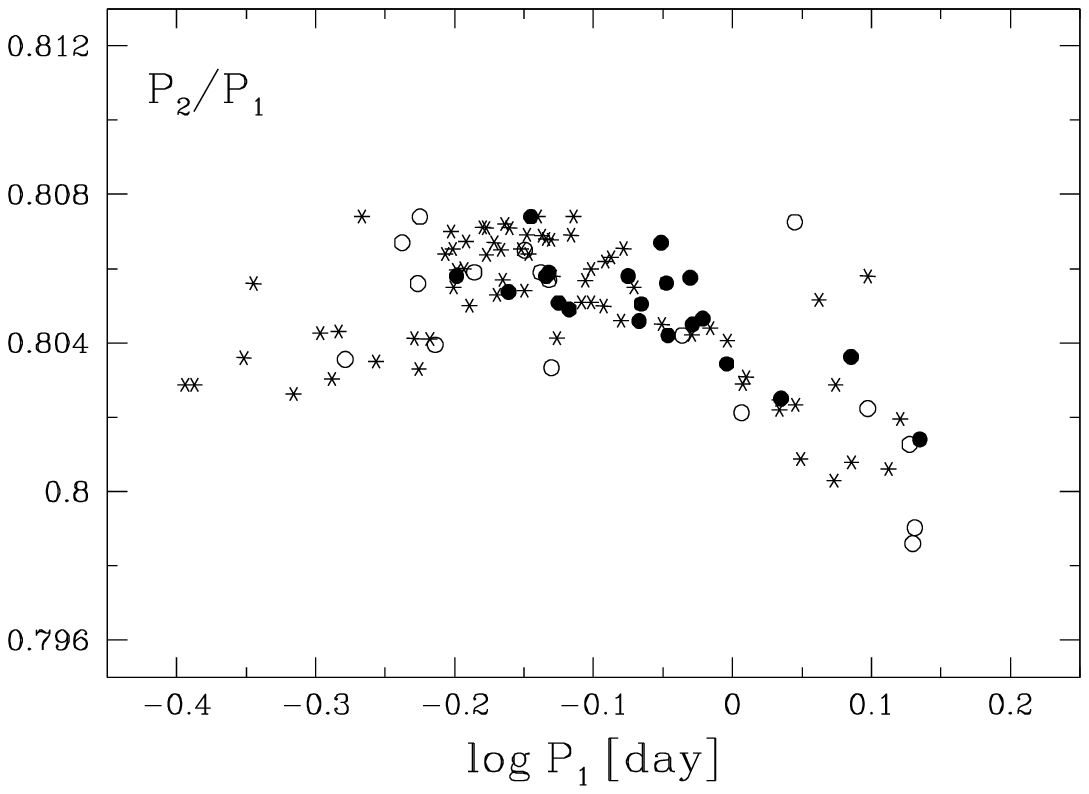}
\vskip -309pt
\FigCap{Petersen diagram for FO/SO double-mode Cepheids in the LMC. Cepheids
        with Blazhko effect and FO/SO-PC variables displayed with filled and
        open circles, respectively. Remaining pulsators plotted with
        asterisks.}
\end{figure}

\begin{figure}%18
\includegraphics[width=18.8cm]{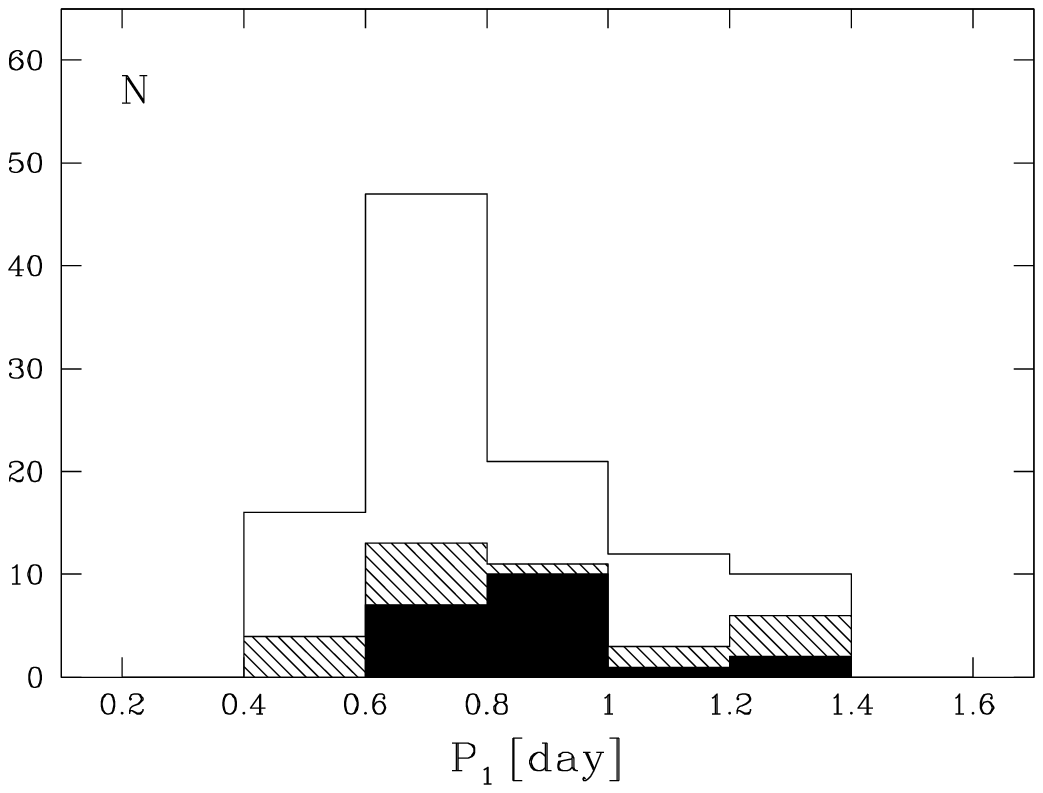}
\vskip -309pt
\FigCap{Distribution of first overtone periods of FO/SO double-mode Cepheids in
        the LMC. Black histogram represents Blazhko pulsators. Shaded area
        marks FO/SO-PC variables.}
\end{figure}

\end{document}